\documentclass[prb,superscriptaddress,twocolumn,floatfix,notitlepage,citeautoscript]{revtex4-2}
\usepackage[T2A]{fontenc}
\usepackage{mmap}
\usepackage{amssymb}
\usepackage{amsmath}
\usepackage{color}
\usepackage[usenames,dvipsnames]{xcolor}
\usepackage{hyperref}
\usepackage{textcomp}
\hypersetup{colorlinks=true,linkcolor=blue,urlcolor=blue,citecolor=blue}
\usepackage[normalem]{ulem}
\usepackage{multirow}
\usepackage{titlesec}
\usepackage{graphicx}
\titleformat{\section}
  {\bf\scshape}{\thesection.}{0.2em}{}

\newcommand{\vor}[1]{{\color{black}#1}}  

\usepackage{array}
\newcolumntype{P}[1]{>{\centering\arraybackslash}p{#1}}

\begin{document}
\title{Path integral approach to bosonisation and nonlinearities in exciton-polariton systems}

\author{Anna M.~Grudinina}
\affiliation{National Research Nuclear University MEPhI (Moscow Engineering Physics Institute), Kashirskoe shosse 31, 115409 Moscow, Russia}
\affiliation{Russian Quantum Center, Skolkovo IC, Bolshoy boulevard 30 bld. 1, 121205 Moscow, Russia}

\author{Nina~S.~Voronova}
\email{nsvoronova@mephi.ru}
\affiliation{National Research Nuclear University MEPhI (Moscow Engineering Physics Institute), Kashirskoe shosse 31, 115409 Moscow, Russia}
\affiliation{Russian Quantum Center, Skolkovo IC, Bolshoy boulevard 30 bld. 1, 121205 Moscow, Russia}

\begin{abstract}
Large exciton-polariton optical nonlinearities present a key mechanism for photonics-based communication, ultimately in the quantum regime. Enhanced nonlinear response from various materials hosting excitons and allowing for their strong coupling with light is therefore the topic of intense studies, both in theoretical and experimental domains. Reports on the scattering rates arising due to various system's nonlinearities, such as the exciton-exciton Coulomb interaction and the Pauli blocking that leads to the saturation of the exciton oscillator strength, however, are contradictory. In this work, we develop a formalism allowing to track the exciton nonlinearities appearing in the regime of strong coupling with photons, that includes finite temperatures, mixing of the exciton excited states, and the dark exciton contributions to saturation self-consistently. The equilibrium path integration approach employed here to address the polariton composite nature, leads to a transparent hierarchy of various contributions to nonlinearity. At the same time, by taking the simplest limit of zero temperature and so-called ``rigid'' excitons, through our framework we retrieve the expressions derived in conventional approaches for exciton interaction constants. In particular, our theory allows to clearly show that such interaction constants cannot be used as fitting parameters tunable in a wide range of values, as they are strictly defined by the material properties, and that other explanations are due for large optical nonlinearities recently reported.
\end{abstract}

\maketitle

\section{Introduction}
\vspace{-10pt}

Exciton-polaritons, hybrid semiconductor quasiparticles that result from strong coupling of electronic excitations with light, are attractive candidates to endow photons with strong nonlinearity, both in the macroscopically coherent regimes and at the single-particle level~\cite{QFL,NatRevQuantum}.
This promising perspective, together with quickly developing state of the art, led in the last years to intensification of experimental and theoretical studies aimed at the analysis of polariton interactions in various material systems, including both conventional semiconductor quantum wells (QWs)~\cite{estrecho, richard} and two-dimensional (2D) transition-metal dichalcogenides (TMDs)~\cite{deng_const, stepanov, menon, zhao, barachati}, the latter allowing to operate at elevated temperatures.

When considering macroscopic phenomena on the polariton level, such as the polariton Bose-Einstein condensation~\cite{BEC_polaritons} and superfluidity~\cite{SF_polaritons}, excitons are most often treated as proper bosons, due to the underlying exciton-photon strong coupling. In the microscopic descriptions on the exciton level, however, it has been argued that their composite electron-hole structure needs to be taken into account. In this context, for electron-hole-photon systems with strong coupling, such as polaritons in a microcavity, several formalisms considering the effective exciton Bose field have been developed, e.g. directly introducing the exciton operators~\cite{yamamoto,parish_ehph1,glazov,combescot}, performing the Usui transformation~\cite{schwendimann}, or basing on the semiconductor Bloch equation~\cite{kira,binder_scat,khurgin1}. All these methods represent generally the same approach describing the bosonisation of interacting fermionic constituents of the exciton in presence of the photon field in the case of low densities ($n_{\rm ex}a^2\ll1$, where $n_{\rm ex}$ and $a$ are the 2D exciton density and Bohr radius, respectively). On the other hand, the opposite limit where the coupling to photons is considered dominant and Coulomb electron-hole interaction absent, was studied within the generalized Dicke model in a series of works by Littlewood and co-authors~\cite{PI1,PI2,PI3,PI4,PI5,PI6} and, later, in Ref.~\cite{parish_ehph3} for excitons bound by light. In both of these limits, however, the considerations were mostly limited to zero temperature and spinless electrons and holes (except for Ref.~\cite{glazov}).

It is perhaps instructive to recall that for bulk semiconductors,
the electron-hole systems without strong coupling to photons were originally studied by Keldysh and Kozlov~\cite{keldysh} in 1960s using the Green's functions technique with additional corrections from multiple electron-hole scattering processes. Later, Kiselev and Babichenko~\cite{kiselev} proposed
the description of such electron-hole fermionic system based on the path integral approach. This theory allows for perturbative derivation of the effective exciton action for a weakly-interacting exciton Bose gas starting from the electron-hole formalism. The exciton interaction constant calculated in~\cite{keldysh} and~\cite{kiselev} is equal to \vor{$g_{\rm ex}^{\rm 3D} = (13\pi/3)\,\hbar^2 a_{\rm 3D}/m$}, which corresponds to $1s$-exciton scattering (here $m$ is the exciton \vor{reduced} mass).
In an analogous approach for quasi-2D excitons in semiconductor QWs, the exciton interaction constant \vor{$g_{\rm ex}^{\rm 2D} = (1-315\pi^2/4096)4\pi\hbar^2/m$} was derived in 1995~\cite{tokatly}. Further works treating QW exciton interaction already in the polariton regime~\cite{yamamoto,schwendimann,binder_scat} reproduced this result and, furthermore, underlined another possible source of the polariton optical nonlinearity: the saturation of exciton oscillator strength.

With the emergence of TMD monolayers as a versatile platform hosting excitons and exciton-polaritons, the search for correct estimates of the interaction constants $g_{\rm ex}$ and $g_{\rm sat}$, characterising the exciton pair interaction and saturation, respectively, was renewed (here and below we drop the superscript ``2D'' for clarity). On the one hand, the small exciton Bohr radii and large binding energies hold promise for robust highly-nonlinear TMD polaritons up to room temperatures. On the other hand, the lowest-lying $s$-states of such excitons exhibit significant deviations from conventional Coulomb-bound hydrogenic model~\cite{berkelbach_prb88,chernikov_prl113}, since the interaction between charges in atomically-thin layers is described by the Rytova-Keldysh potential~\cite{keldysh_int}.
Among theoretical works, Ref.~\cite{parish_pol},  revisiting the logarithmic factor in the interaction strength---a well-known result for 2D boson scattering problem~\cite{LozYudson,Castin_notes}---underlined that for polaritons the interaction constants should be larger than those for excitons, due to the difference between the exciton and lower-polariton (LP) energies. Ref.~\cite{trion-polaritons} in their Supplemental Information provides the detailed derivation of the Rabi splitting renormalisation (saturation) in the operator formalism, similar to~\cite{yamamoto,combescot,glazov}, but up to the second order in $n_{\rm ex}a^2$. Thus, the analytical expressions for the constants $g_{\rm ex}$ and $g_{\rm sat}$ describing both polariton nonlinearities, valid for the case of robust $1s$--excitons at $T=0$, exist.
Nevertheless, recent experiments reveal not only a substantial disagreement with those predictions, but also contradictory results in comparison to each other. In particular, while some giant exciton saturation constants were recently reported for $1s$-excitons in WS$_2$~\cite{menon}, Ref.~\cite{zhao} reports on the numbers two orders of magnitude smaller for the same material. Still, the saturation-related nonlinearity $g_{\rm sat}$ in both works exceeds $g_{\rm ex}$. 
Similar studies performed in~\cite{stepanov} for MoSe$_2$ revealed $g_{\rm ex}$ and $g_{\rm sat}$ to be of the same order, whereas in \cite{deng_const} these constants are smaller by one and two orders of magnitude, respectively. Notably, a very recent experiment performed with GaAs QW polaritons, which was able to resolve the upper-polariton (UP) branch, also revealed $g_{\rm sat}\gg g_{\rm ex}$~\cite{richard}.
\vor{Noting such vast disagreements in the reported values of polariton-polariton nonlinearities, it needs to be noted that from the experimental point of view, the definition of polariton interaction constants is very nontrivial. First complication comes from the intrinsic difficulty of independent measurements of the particle densities. Second, one needs independent observables to distinguish between the interaction and saturation mechanisms of the nonlinearity, which  requires a clear observation of both the upper and lower polariton branches.}
In all the above experimental works, the two constants $g_{\rm ex}$ and $g_{\rm sat}$ are used as free parameters to fit the observed dispersions shifts with \vor{excitation power and, as a result, the obtained values are completely disconnected from the microscopic theoretical descriptions}. All these results call for interpretation and detailed analysis.

In this work, we develop the path integral approach~\cite{kiselev} for an electron-hole-photon mixture, examining the influence of the photon field on the shape of the exciton field, and derive the effective exciton-photon action. While the considerations presented here in their simplest limit ($T=0$ and the rigid $1s$-exciton) do not lead to results other than those obtained in previous works~\cite{yamamoto,glazov, schwendimann,binder_scat}, they differ in regard to their derivation. In its full shape, however, our theory provides means to follow the influence of finite temperatures, electron-electron scattering, and the reshaping of the exciton structure due to the presence of photons (such as forming the ``flexible'' exciton where many $s$-states are mixed) on the exciton nonlinearities.
To provide possible explanation of experimentally-observed saturation constants, the dark exciton states are self-consistently introduced in the exciton-photon action via consideration of the electron and hole spins.

The paper is organized as follows. Section~\ref{sec_action} is devoted to the full derivation of the exciton-photon action at finite temperature in the spinless case. Sec.~\ref{sec_const} addresses the limit $T=0$ and rigid $1s$-exciton state, which allows to compare results with the existing works. In Sec.~\ref{sec_rabi} we present the corrections appearing in the renormalization of the Rabi splitting due to contributions of dark excitons to the UP and LP dispersions shifts when the spins of underlying electron and holes are self-consistently taken into account. Sec.~\ref{sec_concl} concludes our studies. The main text is supplemented with Appendices~\ref{appA}--\ref{appD} containing some cumbersome calculation details\vor{, including the extension of the main theory accounting for spins}.

\section{The exciton-photon action} \label{sec_action}
\vspace{-10pt}
In the presented derivation, we consider the most relevant case $n_{\rm ex} a^2 \ll1$. The opposite limit can be attributed to the formation of the exciton insulator proposed by Keldysh and Kopaev~\cite{keldysh_1}, which for polariton systems was studied in Ref.~\cite{marchetti}. We start with the action for the electron-hole-photon system so far neglecting the spins of particles, in terms of the field operators of electrons $\Psi_{\!c(v)}$ in the conduction (valence) bands with effective masses $m_{c(v)}$ and the dispersions $\varepsilon_{\!c(v)}({\bf k}) = \pm \left(E_{\rm g}/2 + \hbar^2 {\bf k}^2/2m_{c(v)}\right)$, and the field operator $\Psi_{\rm ph}$ of photons in a microcavity, with the effective mass $m_{\rm ph}$ and the dispersion $E_{\rm ph}({\bf k}) = E_{\rm ph}^0 + \hbar^2 {\bf k}^2/2 m_{\rm ph}$. The origin of the energy is taken in the middle of the bandgap of width $E_{\rm g}$, and $E_{\rm ph}^0$ represents the cavity cutoff. The action reads:
\vspace{-10pt}
\begin{multline}\label{action_e-h-ph_initial}
    \mathcal{S}[\Psi_{\!c},\Psi_{\!v}, \Psi_{\rm ph}] = \! \int \!\! d{\bf r} \!\int\limits_{0}^{\beta} \!d\tau \\
    \left[\!\begin{pmatrix}
            \overline{\Psi}_{\!c}(x) & \!\!\overline{\Psi}_{\!v}(x)
        \end{pmatrix}\!\!
        \begin{pmatrix}
            \partial_{\tau} \!+\! \varepsilon_{\!c}({\bf\hat{k}}) \!-\! \mu_{c} & 0 \\ 0 &  \!\!\!\partial_{\tau} \!+\! \varepsilon_{\!v}({\bf\hat{k}}) \!-\! \mu_{v}
        \end{pmatrix} \!\!
        \begin{pmatrix}
             {\Psi}_{\!c}(x) \\ {\Psi}_{\!v}(x)
        \end{pmatrix} \right.  \\
        +  \overline{\Psi}_{\rm ph}(x)(\partial_{\tau} + E_{\rm ph}({\bf\hat{k}}) - \mu_{\rm ph}) \Psi_{\rm ph}(x) \\
        \biggl. + g_{\rm R} \left(\overline{\Psi}_{\rm ph}(x)\Psi_{\!c}(x)\overline{\Psi}_{\!v}(x) +\overline{\Psi}_{\!c}(x)\Psi_{\!v}(x)\Psi_{\rm ph}(x) \right)\biggr] \\[-5pt]
        + \frac{1}{2} \!\sum\limits_{i,j} \!\int \!\! d{\bf r}d{\bf r}^\prime \!\! \int\limits_{0}^{\beta} \! d\tau d\tau^\prime V(x-x^\prime) \overline{\Psi}_{\!i}(x)\Psi_{\!i}(x) \overline{\Psi}_{\!j}(x^\prime) \Psi_{\!j}(x^\prime),
\end{multline}
with $i,j$ running over $\{c,v\}$, $x = ({\bf r}, \tau)$, $\tau$ is the imaginary time, $\beta=\hbar/k_{\rm B}T$, ${\bf\hat{k}}$ is the momentum operator, and $V(x-x^\prime) \equiv V({\bf r-r^\prime})\delta(\tau-\tau^\prime)$, where $V({\bf r-r^\prime})$ represents the potential of interaction between charged particles. $\mu_{c(v)}$ is the chemical potential of the electrons in conduction (valence) band (which approximately equals zero for the ground state in undoped semiconductors), and $\mu_{\rm ph}$ is the chemical potential of photons. We consider separate chemical potentials of species since the total number of particles (electrons, holes, and photons) is not conserved. Finally, $g_{\rm R}$ is the amplitude of light-matter coupling (electron-hole annihilation with photon creation and vice versa). \vor{The inclusion of the particles spins and valleys into Eq.~(\ref{action_e-h-ph_initial}) is discussed below.}
In the functional integration (path integration) approach the grand canonical partition function is defined as:
\begin{equation}
    \mathcal{Z} =\int \mathcal{D}[\Psi_{\!c,v}, \overline{\Psi}_{\!c,v}]\mathcal{D}[\Psi_{\rm ph}, \overline{\Psi}_{\rm ph}]e^{-\mathcal{S}[\Psi_{\!c},\Psi_{\!v}, \Psi_{\rm ph}]}\,.
\end{equation}
The exciton field $\Delta({\bf r}_1,{\bf r}_2; \tau) = \overline{\Psi}_{\!v}({\bf r}_2,\tau)\Psi_{\!c}({\bf r}_1,\tau)$ can be introduced in the following way:
\begin{align}
    \mathcal{Z} = &\int \!\mathcal{D}[\Psi] e^{-\mathcal{S}[\Psi]} \\[-5pt]
    = &\int \! \mathcal{D}[\Psi] \mathcal{D}[\bar{\Delta}\Delta] e^{-\mathcal{S}[\Psi]}
    \delta(\Delta - \overline{\Psi}_{\!v}\Psi_{\!c}) \delta(\bar{\Delta} - \overline{\Psi}_{\!c}\Psi_{\!v}) \nonumber\\
    = &\int \!\mathcal{D}[\Psi] \mathcal{D}[\bar{\Delta}\Delta] \mathcal{D}[\bar{\phi},\phi]e^{-\mathcal{S}[\Psi]}e^{i\bar{\phi}(\Delta - \overline{\Psi}_{\!v}\Psi_{\!c}) + i\phi (\bar{\Delta} - \overline{\Psi}_{\!c} \Psi_{\!v})}, \nonumber
\end{align}
where $\phi({\bf r}_1,{\bf r}_2; \tau)$ is an auxiliary field. Looking ahead, this field is unphysical and below will be integrated over.
Strictly speaking, other fields can be taken into account using the standard Hubbard-Stratonovich transformation: one can consider different channels of dual-field decoupling in the interacting part of the initial action. In particular, introducing the dual field $\overline{\Psi}_i(x)\Psi_{i}(y)$ ($i=c,v$), one can derive the renormalization of the exciton-exciton interaction matrix element due to electron scattering, which is done in Appendix~\ref{appA}.

For convenience we perform the Fourier transform assuming the system homogeneous and finite, and account for finite temperatures:
\begin{eqnarray*}
     \Psi_{\!i}(x,\tau) &=& \sqrt{\frac{T}{S}}\sum_{\bf k}\,\sum_{\omega_{n}}\Psi_{\!i}({\bf k}, \omega_{n}) e^{i {\bf k}\cdot{\bf x} - i\omega_{n}\tau},\nonumber\\[-2pt]
    \Delta(x, y) &=& \frac{\sqrt{T}}{S}\sum_{{\bf k},{\bf k}^\prime} \sum_{\Omega_{n}} \Delta({\bf k}, {\bf k}^\prime\!, \Omega_n) e^{i{\bf k}\cdot{\bf x} - i {\bf k}^\prime\!\cdot{\bf y} - i\Omega_{n} \tau},\nonumber\\[-2pt]
   \phi(x, y) &=& \frac{\sqrt{T}}{S}\sum_{{\bf k},{\bf k}^\prime} \sum_{\Omega_{n}} \phi({\bf k}, {\bf k}^\prime\!, \Omega_n) e^{i{\bf k}\cdot{\bf x} - i {\bf k}^\prime\!\cdot{\bf y} - i\Omega_{n} \tau},
\end{eqnarray*}
where $\omega_{n}=(2\pi n+\pi)/\beta$ and $\Omega_n=2\pi n/\beta$ are the fermionic and bosonic Mastubara frequencies, respectively, $n\in\mathbb{Z}$, ${\bf k},{\bf k}^\prime$ are the quantized wavevectors, and $S$ is the system area. The action takes the following form:
\begin{widetext}
\begin{multline}
    \mathcal{S}[\Psi_{\!c},\Psi_{\!v},\Psi_{\rm ph}, \phi, \Delta]=\sum_{{\bf k},\omega}\left[
        \begin{pmatrix}
            \overline{\Psi}_{\!c}(k) & \overline{\Psi}_{\!v}(k)
        \end{pmatrix}\!\!
        \begin{pmatrix}
            -i \omega + \varepsilon_c({\bf k}) - \mu_{c} & 0 \\ 0 &   -i \omega + \varepsilon_v({\bf k}) - \mu_{v}
        \end{pmatrix}\!\!
        \begin{pmatrix}
             {\Psi}_{\!c}(k) \\ {\Psi}_{\!v}(k)
        \end{pmatrix}
    \right] \\[3pt]
    + i\sqrt{T} \sum_{{\bf k}_{1},\omega_{1}}\sum_{{\bf k}_{2},\omega_{2}} \!\!
    \left[
        \begin{pmatrix}
            \overline{\Psi}_{\!c}(k_1) & \overline{\Psi}_{\!v}(k_1)
        \end{pmatrix}\!\!
        \begin{pmatrix}
           0 & \phi({\bf k}_{1},{\bf k}_{2}, \omega_{1} \!-\! \omega_{2})\\
           \bar{\phi}({\bf k}_{2},{\bf k}_{1}, \omega_{2} \!-\! \omega_{1}) & 0
        \end{pmatrix}\!\!
        \begin{pmatrix}
             {\Psi}_{\!c}(k_2) \\ {\Psi}_{\!v}(k_2)
        \end{pmatrix} \right] \\[3pt]
    - i \sum_{{\bf k}_{1},{\bf k}_{2},\Omega} \left[ \bar{\phi}({\bf k}_{1},{\bf k}_{2}, \Omega)\Delta({\bf k}_{1},{\bf k}_{2}, \Omega) + \bar{\Delta}({\bf k}_{1},{\bf k}_{2}, \Omega) \phi({\bf k}_{1},{\bf k}_{2}, \Omega) \right]
    \\[3pt]
    - \!\!\!\!\sum_{{\bf k}_{1}\dots{\bf k}_{4}, \Omega} \!\!\! V({\bf k}_{1} \!-\! {\bf k}_{2}) \bar{\Delta}({\bf k}_{1},{\bf k}_{4},\Omega) \Delta({\bf k}_2,{\bf k}_3,\Omega) \delta({\bf k}_1 \!+\! {\bf k}_3, {\bf k}_2 \!+\! {\bf k}_4) + \sum_{{\bf k}, \Omega} \overline{\Psi}_{\rm ph}(k)(- i \Omega + E_{\rm ph}({\bf k}) - \mu_{\rm ph})\Psi_{\rm ph}(k) \\[3pt]
    + \frac{g_{\rm R}}{\sqrt{S}} \sum_{{\bf k}_1,{\bf k}_2,\Omega} \left[\overline{\Psi}_{\rm ph}({\bf k}_1 \!-\! {\bf k}_2,\Omega) \Delta({\bf k}_1,{\bf k}_2,\Omega) + \bar{\Delta}({\bf k}_1,{\bf k}_2,\Omega)\Psi_{\rm ph}({\bf k}_1 \!-\! {\bf k}_2,\Omega)\right],
    \label{action1}
\end{multline}
\end{widetext}
where all sums over $\omega,\Omega$ involve summation over integer indices $n$ (subscripts $n$ are omitted for clarity), and the four-vector notation $k \equiv ({\bf k}, \omega_n)$ or $({\bf k},\Omega_n)$ was introduced for both fermionic and bosonic fields.
It is then straightforward to perform the integration over the fermionic fields $\Psi_{\!c,v}$, which results in the $(-{\rm Trln}\mathcal{G}^{-1})$ contribution to the action, where $\mathcal{G}^{-1} = \mathcal{G}^{-1}_{0}+ \delta \mathcal{G}^{-1}$ with 
$$\mathcal{G}^{-1}_{0}\!=\!\begin{pmatrix}
    -i \omega_1 \!+\! \varepsilon_c({\bf k}_1) \!-\! \mu_{c} & 0 \\ 0 &   -i \omega_1 \!+\! \varepsilon_v({\bf k}_1) \!-\! \mu_{v}
    \end{pmatrix} \delta_{k_1k_2}$$
denoting the bare electron Green's function, \vor{$\delta_{k_1,k_2}$ being the Kronecker symbol in four-vector notation,} and 
$$\delta\mathcal{G}^{-1} \!= i\sqrt{T}
       \begin{pmatrix}
           0 & \phi({\bf k}_1,{\bf k}_2, 
           \Omega) \\\bar{\phi}({\bf k}_2,{\bf k}_1, 
           -\Omega) & 0
        \end{pmatrix}$$
containing the introduced auxiliary field $\phi$ (here the bosonic Matsubara frequency $\Omega\equiv\Omega_n$ originates from the difference of two fermionic frequencies $\omega_{1n} - \omega_{2n}$). It can be easily calculated under the assumption $\delta \mathcal{G}^{-1} \!\!\ll \mathcal{G}^{-1}_{0}$ (considering $\delta \mathcal{G}^{-1}$ as a perturbation, which corresponds to our applicability condition $n_{\rm ex} a^2 \ll1$):
\begin{multline}
    {\rm Trln}\mathcal{G}^{-1} = {\rm Trln}\mathcal{G}_{0}^{-1} + {\rm Trln}(1+\mathcal{G}_0\delta\mathcal{G}^{-1}) =\\=  {\rm Trln}\mathcal{G}_{0}^{-1} + {\rm Tr} \sum_{n} \frac{(-1)^{n-1}}{n}(\mathcal{G}_0\delta\mathcal{G}^{-1})^n.
\end{multline}
In the expansion series one should keep the terms up to the 4th power to derive the exciton-exciton interaction.

Since $\mathcal{G}_{0}$ is diagonal while $\delta\mathcal{G}^{-1}$ is antidiagonal, the first and third terms in the logarithm expansion are absent, hence in the lowest order in the auxiliary field $\phi$ one gets:
\begin{align}
    \frac{1}{2}{\rm Tr}(\mathcal{G}_0\delta\mathcal{G}^{-1})^2 = & -T \!\!\sum_{{\bf k}_1, {\bf k}_2, \Omega} \!\bar{\phi}({\bf k}_1,{\bf k}_2, \Omega)\phi({\bf k}_1,{\bf k}_2, \Omega) \nonumber \\
    &\times \!\sum_{\omega^\prime} \mathcal{G}_{0}({\bf k}_1, \omega^\prime \!+\! \Omega)_{11} \mathcal{G}_{0}({\bf k}_2, \omega')_{22}.
\end{align}
The last summation is performed over the fermionic Matsubara frequency $\omega^\prime$ of the loop diagram, and for finite temperatures it leads to the exciton propagator
\begin{multline}\label{A7}
  T \sum_{\omega^\prime} \mathcal{G}_{0}({\bf k}_1, \omega^\prime + \Omega)_{11} \mathcal{G}_{0}({\bf k}_2, \omega^\prime)_{22} = \\
  =\frac{n_{v}({\bf k}_2)-n_{c}({\bf k}_1)}{\varepsilon_c({\bf k}_1) \!-\! \varepsilon_v({\bf k}_2) \!-\! \mu_{c} \!+\! \mu_{v} \!-\! i\Omega} \equiv \mathcal{A}_{{\bf k}_1{\bf k}_2}^{\,\,\Omega},
\end{multline}
where $n_{i}({\bf k}) = 1/(e^{\beta (\varepsilon_{i}({\bf k}) -\mu_{i})}+1)$ is the Fermi distribution function in the $i$-band  (the details are provided in Appendix \ref{appE}). Here we note that for an undoped semiconductor $n_{c} = n_{v}$ and thus one can approximate $\mu_{c}\sim -\mu_{v} + T\ln(m_{v}/m_{c})$. From the point of view of equilibrium chemical reactions, the exciton chemical potential is defined as $\mu_{\rm ex} = \mu_{e} + \mu_{h} = \mu_{c} - \mu_{v}$ (where $\mu_{e,h}$ denote the chemical potentials of the electrons and holes), and in thermodynamic equilibrium between the excitons and photons
it should equal $\mu_{\rm ex} = \mu_{\rm ph}\equiv \mu$. Then the above expression reduces to
\begin{equation}\label{Ak1k2}
     \mathcal{A}_{{\bf k}_1{\bf k}_2}^{\,\,\Omega} = \frac{ n_{v}({\bf k}_2)-n_{c}({\bf k}_1)}{\varepsilon_c({\bf k}_1) - \varepsilon_v({\bf k}_2) -\mu-i \Omega}.
\end{equation}

In the same manner we calculate the $|\phi|^4$-term:
\begin{multline}\label{phi4}
    \frac{1}{4}{\rm Tr}(\mathcal{G}_0\delta\mathcal{G}^{-1})^4 = \frac{2}{4}(i\sqrt{T})^4\!\!\sum_{{\bf k}_1\dots{\bf k}_4}\sum_{\Omega_1\dots\Omega_3} \!\! \phi({\bf k}_1,{\bf k}_2, \Omega_1)\\[5pt]
    \times\bar{\phi}({\bf k}_3, {\bf k}_2, \Omega_2)\phi({\bf k}_3, {\bf k}_4, \Omega_3)\bar{\phi}({\bf k}_1, {\bf k}_4, \Omega_1 \!+\! \Omega_3 \!-\! \Omega_2)
    \\[3pt]
    \times \sum_{\omega^\prime}
    \mathcal{G}_{0}({\bf k}_1, \omega^\prime)_{11}\mathcal{G}_0({\bf k}_2, \omega^\prime \!-\! \Omega_1)_{22}\mathcal{G}_0({\bf k}_3, \omega^\prime \!-\! \Omega_1 \!+\! \Omega_2)_{11} \\ \times\mathcal{G}_0({\bf k}_4, \omega^\prime- \Omega_1+ \Omega_2- \Omega_3)_{22},
\end{multline}
where the factor of 2 arises from the number of loops in the corresponding diagram series (see Appendix~\ref{appE} and Fig.~\ref{fig:loops} therein).
The result of summation over the frequency of the fourth-order loop diagram in Eq.~(\ref{phi4}) is given in Appendix~\ref{appE}, while here we only introduce the loop notation:
\begin{multline}
   \mathcal{L}_{{\bf k}_1{\bf k}_2{\bf k}_3{\bf k}_4}^{\Omega_1\Omega_2\Omega_3} = T \sum_{\omega^\prime}
    \mathcal{G}_{0}({\bf k}_1, \omega^\prime)_{11}\mathcal{G}_0({\bf k}_2, \omega^\prime \!\!-\! \Omega_1)_{22} \\
    \times \mathcal{G}_0({\bf k}_3, \omega^\prime \!\!-\! \Omega_1 \!+\! \Omega_2)_{11} \mathcal{G}_0({\bf k}_4, \omega^\prime \!\! -\! \Omega_1 \!+\! \Omega_2 \!-\! \Omega_3)_{22}
\end{multline}
The effective action takes the form:
\begin{widetext}
\begin{multline}
     \mathcal{S}[\Psi_{\rm ph},\phi, \Delta]= - \!\!\sum_{{\bf k}_1, {\bf k}_2, \Omega} \!\! \mathcal{A}_{{\bf k}_1{\bf k}_2}^{\,\,\Omega} \, \bar{\phi}({\bf k}_1,{\bf k}_2, \Omega)\phi({\bf k}_1,{\bf k}_2, \Omega) + \sum_{{\bf k}, \Omega} \overline{\Psi}_{\rm ph}(k)(- i \Omega + E_{\rm ph}({\bf k}) -\mu)\Psi_{\rm ph}(k)
     \\
     -i \!\!\!\sum_{{\bf k}_1,{\bf k}_2,\Omega} \!\!\!\bigl[\bar{\phi}({\bf k}_1,{\bf k}_2, \Omega)\Delta({\bf k}_1,{\bf k}_2, \Omega) + \bar{\Delta}({\bf k}_1,{\bf k}_2, \Omega) \phi({\bf k}_1,{\bf k}_2, \Omega) \bigr]
     - \!\!\!\!\! \sum_{{\bf k}_1\dots{\bf k}_4, \Omega} \!\!\!\!\! V({\bf k}_1 - {\bf k}_2) \bar{\Delta}({\bf k}_1,{\bf k}_4,\Omega) \Delta({\bf k}_2,{\bf k}_3,\Omega) \delta({\bf k}_1+{\bf k}_3, {\bf k}_2+{\bf k}_4)
     \\
     + \frac{g_{\rm R}}{\sqrt{S}} \sum_{{\bf k}_1,{\bf k}_2,\Omega} \left[\overline{\Psi}_{\rm ph}({\bf k}_1-{\bf k}_2,\Omega) \Delta({\bf k}_1,{\bf k}_2,\Omega) + \bar{\Delta}({\bf k}_1,{\bf k}_2,\Omega)\Psi_{\rm ph}({\bf k}_1-{\bf k}_2,\Omega)\right]
     \\
     +\frac{T}{2}\sum_{\{{\bf k}_i, \Omega_i\}} \!\!\! \mathcal{L}_{{\bf k}_1{\bf k}_2{\bf k}_3{\bf k}_4}^{\Omega_1\Omega_2\Omega_3} \, \phi({\bf k}_1,{\bf k}_2, \Omega_1)\bar{\phi}({\bf k}_3, {\bf k}_2, \Omega_2)\phi({\bf k}_3, {\bf k}_4, \Omega_3)\bar{\phi}({\bf k}_1, {\bf k}_4, \Omega_1 + \Omega_3 -\Omega_2).
\end{multline}
\end{widetext}

Now, the auxiliary field $\phi$ needs to be excluded. Within the saddle-point approximation in the lowest order of the ${\rm Trln}\mathcal{G}^{-1}$ expansion we have:
\begin{equation}
    \frac{\delta \mathcal{S}}{\delta \bar{\phi}} \!=\! - \mathcal{A}_{{\bf k}_1{\bf k}_2}^{\,\,\Omega}\, \phi({\bf k}_1,{\bf k}_2,\Omega) - i \Delta({\bf k}_1,{\bf k}_2,\Omega) =0,
\end{equation}
and the relation between the exciton and the auxiliary fields can be found:
\begin{subequations}
    \begin{eqnarray}
    \phi({\bf k}_1,{\bf k}_2,\Omega) = \frac{ \Delta({\bf k}_1,{\bf k}_2,\Omega)}{i\mathcal{A}_{{\bf k}_1{\bf k}_2}^{\,\,\Omega}}, \\
    \bar{\phi}({\bf k}_1,{\bf k}_2,\Omega) =\frac{\bar{\Delta}({\bf k}_1,{\bf k}_2,\Omega)}{i\mathcal{A}_{{\bf k}_1{\bf k}_2}^{\,\,\Omega}},
    \end{eqnarray}
\end{subequations}
which allows one to rewrite the effective action in terms of the exciton and photon fields, and obtain the effective exciton-photon action. Despite the result being quite similar with the previously reported~\cite{lozovik_PI}, in our case the exciton field depends on two momenta. We can interpret it as the exciton field $\Delta({\bf k}_1,{\bf k}_2, \Omega)$ corresponding to the relative motion of electrons belonging to different bands, with the momentum ${\bf p} = (m_{c}{\bf k}_1+m_{v}{\bf k}_2)/(m_{c}+ m_{v})$, and motion of the exciton as a whole, with the momentum ${\bf k}={\bf k}_1-{\bf k}_2$.

\subsection*{Bethe-Salpeter equation for excitons coupled to photons at finite temperature}
\vspace{-10pt}

It is useful to rewrite the obtained expressions in terms of the relative and total momenta:
\begin{multline}
      \mathcal{S}[\Psi_{\rm ph},\Delta]= \sum_{{\bf p}, {\bf k}, \Omega} \frac{\bar{\Delta}({\bf p},{\bf k}, \Omega)\Delta({\bf p},{\bf k}, \Omega)}{\mathcal{A}_{{\bf p}{\bf k}}^{\,\Omega}} \\
      + \sum_{{\bf k}, \Omega} \overline{\Psi}_{\rm ph}({\bf k},\Omega)(- i \Omega + E_{\rm ph}({\bf k}) -\mu)\Psi_{\rm ph}({\bf k},\Omega) \\
      - \sum_{{\bf k},{\bf p}, {\bf q}, \Omega} V({\bf p} -{\bf q}) \bar{\Delta}({\bf p},{\bf k},\Omega) \Delta({\bf q},{\bf k},\Omega)  \\
      +\! \frac{g_{\rm R}}{\sqrt{S}} \!\! \sum_{{\bf p},{\bf k},\Omega} \!\! \!\left[\overline{\Psi}_{\rm ph}({\bf k},\Omega) \Delta({\bf p},{\bf k},\Omega) \!+\! \bar{\Delta}({\bf p},{\bf k},\Omega)\Psi_{\rm ph}({\bf k},\Omega)\right]\\
      + \frac{T}{2} \!\! \sum_{{\bf p},\{{\bf l}_i, \Omega_i\}} \!\!\!\! \tilde{\mathcal{L}}_{{\bf p}\,{\bf l}_1{\bf l}_2{\bf l}_3}^{\Omega_1 \Omega_2\Omega_3} \, \Delta({\bf p},{\bf l}_1, \Omega_1)\bar{\Delta}({\bf p} +\tfrac{{\bf l}_2 \!-\! {\bf l}_1}{2}\!, {\bf l}_2, \Omega_2) \\
      \times \!\Delta({\bf p} +{\bf l}_2\!- \tfrac{{\bf l}_1 \!-\! {\bf l}_3}{2}\!, {\bf l}_3, \Omega_3)\bar{\Delta}({\bf p} + \tfrac{{\bf l}_2 \!-\! {\bf l}_3}{2}\!, {\bf l}_1 - {\bf l}_2+ {\bf l}_3, \Omega_1 + \Omega_3 - \Omega_2).
      \nonumber
\end{multline}
Here and below, tildes denote the loops connecting exciton fields $\Delta$ (see Appendix~\ref{appE}). When performing the saddle-point approximation to define the shape of the exciton field in presence of photons, the last term can be considered as perturbation since it was derived in the higher order in the expansion series, so that 
it can be temporarily neglected:
\begin{subequations}
\begin{eqnarray}
    \frac{\delta \mathcal{S}}{\delta\bar{\Delta}} &=&  \frac{\Delta({\bf p},{\bf k}, \Omega)}{\mathcal{A}_{{\bf p}{\bf k}}^{\,\Omega}} + \frac{g_{\rm R}}{\sqrt{S}}\Psi_{\rm ph}({\bf k}, \Omega) \nonumber\\
    & & \quad - \sum_{{\bf q}}V({\bf p} -{\bf q}) \Delta( {\bf q},{\bf k},\Omega)=0, \label{saddle-point-eqA} \\
     \frac{\delta \mathcal{S}}{\delta\overline{\Psi}_{\rm ph}} &=& (- i \Omega + E_{\rm ph}({\bf k}) -\mu)\Psi_{\rm ph}({\bf k},\Omega) \nonumber \\
    & & \qquad\quad + \sum_{{\bf p}}\frac{g_{\rm R}}{\sqrt{S}}\Delta({\bf p},{\bf k}, \Omega)=0. \label{saddle-point-eqB}
\end{eqnarray}
\end{subequations}
Determining the photon field via the exciton field in (\ref{saddle-point-eqB}) and substituting it into (\ref{saddle-point-eqA}), we obtain the generalized Bethe-Salpeter equation modified by the presence of the photon field (similar to the semiconductor Bloch equation~\cite{kira} albeit with the explicit dependence on ${\bf k}$ in the photon dispersion):
\begin{multline}\label{BSE}
    \frac{\Delta({\bf p},{\bf k}, \Omega)}{\mathcal{A}_{{\bf p}{\bf k}}^{\,\Omega}} - \sum_{{\bf q}}\Biggl[V({\bf p} -{\bf q}) \Biggr. \\
    + \left. \left(\frac{g_{\rm R}}{\sqrt{S}}\right)^{\!\!2}\!\frac{1}{E_{\rm ph}({\bf k}) -\mu- i \Omega} \right]\Delta( {\bf q},{\bf k},\Omega)=0.
\end{multline}

This equation is one of the main results of this study. Finite temperatures are contained in the summations and Fermi distributions appearing in $\mathcal{A}_{{\bf p}{\bf k}}^{\,\Omega}$. As anticipated, the photon field alters the exciton interaction, since the electrons and holes in such a system can be bound not only by Coulomb-like interactions, but also by light. This phenomenon is known in the literature as the flexible exciton limit~\cite{khurgin} in which the polariton Rabi splitting is not negligible compared to the exciton binding energy: $\hbar \Omega_{\rm R}\sim E_{\rm b}$.
It should be noted that similar equations were recently derived in Ref.~\cite{parish_ehph1,parish_ehph3} for the zero-temperature case using the operator formalism. Ref.~\cite{khurgin1} proposes the solution of (\ref{BSE}) for $T=0$ in the case of one-mode fields. To compare, we apply this method to the equation (\ref{BSE}) at zero temperature in Appendix~\ref{appB}.

Eq.~(\ref{BSE}) can be solved using the ansatz $$\Delta({\bf p}, {\bf k},\Omega) = \sum_{\nu}\chi^{(\nu)}_{{\bf k}, \Omega}({\bf p})C^{(\nu)}({\bf k}, \Omega),$$
which is a decomposition of the exciton field over the basis of (hydrogen-like) wavefunctions $\chi^{(\nu)}_{{\bf k},\Omega}({\bf p})$, with $\nu$ being the number of excitonic $\nu s$--state and $C^{(\nu)}({\bf k}, \Omega)$ the corresponding exciton field depending only on the total momentum. Introducing the interaction matrix element $g({\bf l}_1,{\bf l}_2,{\bf l}_3, \Omega_1,\Omega_2,\Omega_3,\nu_1,\nu_2,\nu_3,\nu_4)\equiv g({\bf l}_i, \Omega_i, \nu_i)$ as
\begin{multline}
    g({\bf l}_i, \Omega_i, \nu_i) \!=\! \sum_{{\bf p}}
    \tilde{\mathcal{L}}_{{\bf p}\,{\bf l}_1{\bf l}_2{\bf l}_3}^{\Omega_1 \Omega_2\Omega_3} \,
    \chi^{(\nu_1)}_{{\bf l}_1, \Omega_1}\!({\bf p})\bar{\chi}^{(\nu_2)}_{{\bf l}_2, \Omega_2}\!({\bf p} +\tfrac{{\bf l_2 - l_1}}{2}\!)
    \\ \qquad\times\chi^{(\nu_3)}_{{\bf l}_3, \Omega_3}({\bf p} +{\bf l}_2- \tfrac{{\bf l_1 - l_3}}{2}\!)\bar{\chi}^{(\nu_4)}_{{\bf l}_1- {\bf l}_2+ {\bf l}_3, \Omega_1-\Omega_2+ \Omega_3}({\bf p} + \tfrac{{\bf l_2 - l_3}}{2}\!),
\end{multline}
we arrive at the final expression for effective exciton-photon action:
\begin{multline}
     \mathcal{S}[\Psi_{\rm ph}, C]= \sum_{{\bf k}, \Omega} \sum_{\nu} \overline{C}^{(\nu)}({\bf k}, \Omega)E^{(\nu)}_{\rm ex}({\bf k}, \Omega) C^{(\nu)}({\bf k}, \Omega)
     \\+ \sum_{{\bf k}, \Omega} \overline{\Psi}_{\rm ph}({\bf k},\Omega)(- i \Omega + E_{\rm ph}({\bf k}) -\mu)\Psi_{\rm ph}({\bf k},\Omega) \\
     + g_{\rm R} \! \sum_{{\bf k}, \Omega} \! \sum_{\nu} \!\left[\bar{\chi}^{(\nu)}_{{\bf k}, \Omega}({\bf r} \!=\! 0) \overline{C}^{(\nu)}\!({\bf k}, \Omega)\Psi_{\rm ph}({\bf k}, \Omega)
     + \text{c.c.} \right]\nonumber 
\end{multline}
\begin{multline}
    + \frac{T}{2}\sum_{\{{\bf l}_i, \Omega_i, \nu_i\}} g({\bf l}_i, \Omega_i, \nu_i)C^{(\nu_1)}({\bf l}_1, \Omega_1)\overline{C}^{(\nu_2)}( {\bf l}_2, \Omega_2)\\\times C^{(\nu_3)}({\bf l}_3, \Omega_3)\overline{C}^{(\nu_4)}( {\bf l}_1- {\bf l}_2+ {\bf l}_3, \Omega_1 + \Omega_3 -\Omega_2).
\label{action-ex-ph-final}
\end{multline}
Here it is important to note that all the expressions are derived in the most general form (for $T\ne 0$, arbitrary interactions and for excitons and photons with ${\bf k}\ne 0$). Summation over $\nu$ allows to account for the flexible exciton case when the coupling to light results in the mixing of states in the exciton $s$-series. Appendix \ref{appB} is devoted to simplifications of these general formulae for the case of zero temperature.

\subsection*{Existing limits}
\vspace{-10pt}
Standard treatment of exciton-polariton systems often neglects
the second term in the square brackets in Eq.~(\ref{BSE}) or, analogously, neglects the terms $\sim g_{\rm R}$ compared to the exciton binding energy $E_{\rm b}$ in Eqs.~(\ref{saddle-point-eqA},\ref{saddle-point-eqB}). Such considerations correspond to the rigid exciton limit when $E_{\rm b}\gg \hbar \Omega_{\rm R}$:
\begin{equation}\label{Wannier_rigid}
    \frac{1}{\mathcal{A}_{{\bf p}{\bf k}}^{\,\Omega}}\Delta({\bf p},{\bf k}, \Omega) - \sum_{{\bf q}}V({\bf p} -{\bf q}) \Delta( {\bf q},{\bf k},\Omega)=0.
\end{equation}
From this point of view, considering only the $1s$-state at $T=0$  and separating variables $\Delta({\bf p}, {\bf k},\Omega) = \chi({\bf p})C({\bf k},\Omega)$ (where $\chi({\bf p})\equiv\chi^{(1)}_{{\bf k},\Omega}({\bf p})$ is independent of ${\bf k}$ and $\Omega$) brings Eq.~(\ref{Wannier_rigid}) to the standard Wannier equation. In this limit, one would obtain the exciton wavefunction and the renormalization of the exciton-photon conversion term previously reported in several works~\cite{schwendimann,parish_ehph1}:
\begin{equation}\label{renorm}
\frac{g_{\rm R}}{\sqrt{S}} \sum_{{\bf q}}\chi({\bf q}) = g_{\rm R} \chi({\bf r} = 0) = \frac{\hbar \Omega_{\rm R}}{2},
\end{equation}
where $\hbar \Omega_{\rm R}/2$ is the experimentally-observed polariton Rabi-splitting. In this case, the saddle-point equations (\ref{saddle-point-eqA}, \ref{saddle-point-eqB}) become the standard Hopfield equations while the exciton-photon action takes the well-known form \cite{lozovik_PI}. Here we emphasize that this approach to treat the action is, in principle, equivalent to the assumption that first an electron and a hole are bound by electrostatic interaction forming an exciton, and then excitons are coupled to photons, which is a common way to describe strong coupling between excitons and photons. Strictly speaking, in the case of finite temperatures variables cannot be divided, as can be anticipated from the shape of $\mathcal{A}_{{\bf p}{\bf k}}^{\,\Omega}$ in Eq.~(\ref{Ak1k2}) with $({\bf k}_1,{\bf k}_2)\rightarrow({\bf p},{\bf k})$.

On the other hand, in the opposite limit of dominating photon-electron-hole coupling, Eq.~(\ref{BSE}) for the exciton field $\Delta$ is similar to~(\ref{Wannier_rigid}) but with the electron-hole interactions induced only by light~\cite{PI6,PI2,PI5,parish_ehph3}, which means one cannot parametrize the field $\Delta$ as $\chi({\bf p})C({\bf k}, \Omega)$ with the hydrogen-like wavefunction $\chi({\bf p})$.

\section{Interaction constants} \label{sec_const}
\vspace{-10pt}
To describe the nonlinearities in terms of interaction constants including the so-called {\it saturation} (exciton-assisted photon-exciton coupling) that arises in our treatment from the fourth-order expansion term, one needs not only to set $T=0$ but also to assume the rigid $1s$-exciton limit. While for the flexible exciton where all $\nu s$ states of the hydrogenic basis get mixed, the saturation contribution is also derived (see Appendix~\ref{appB}), it cannot be presented as a closed-form explicit expression. The case of rigid excitons, in which we can analyze the results of recent experiments~\cite{menon,stepanov,richard,zhao,deng_const} and compare to existing theoretical results~\cite{yamamoto,schwendimann,glazov,trion-polaritons,parish_ehph1}, is considered in Appendix~\ref{appC}.

For the ground-state exciton wavefunction which in momentum space reads $\chi({\bf p}) = 2 a\sqrt{2\pi}/(1 + p^2 a^2)^{3/2}$, the expression for the saturation interaction constant is defined as $g_{\rm sat} = (8 \pi/7) a^2\hbar \Omega_{\rm R}/2$, where $a$ in the hydrogenic description is the 2D exciton Bohr radius, while for the Rytova-Keldysh $1s$-state should be considered a variational parameter~\cite{Shahnazaryan2017, numericalWF}. This is a well-established result obtained in many different approaches~\cite{schwendimann,glazov,trion-polaritons}. Clearly, the exciton-assisted exciton-photon coupling $\sim g_{\rm sat}n_{\rm ex}$ describes the conversion between photons and excitons in the higher order of the small parameter $n_{\rm ex}a^2$ compared to the terms $\sim(\overline{\Psi}_{\rm ph} C + {\rm c.c.})$.
The corresponding reduction of the polariton Rabi splitting can be defined from the shifts of the UP and LP dispersions with the increase of the exciton density. Performing the saddle-point approximation for excitons and photons in the assumption of uniform exciton density $n_{\rm ex} = \text{const}$ (see Appendix \ref{appC}), we obtain 
the splitting between the dispersions minima renormalized by interactions 
\begin{eqnarray}
    \hbar\Omega_{R}(n_{\rm ex})&=& \hbar\Omega_{\rm R}\sqrt{1  - 4 \frac{g_{\rm sat}n_{\rm ex}}{\hbar \Omega_{\rm R}} + \frac{3 (g_{\rm sat}n_{\rm ex})^2 }{(\hbar \Omega_{R})^2}}
   \nonumber \\
   & \approx & \hbar\Omega_{\rm R} - 2g_{\rm sat}n_{\rm ex}.
\end{eqnarray}
Since $g_{\rm sat}n_{\rm ex} \sim \hbar\Omega_{\rm R} n_{\rm ex}a^{2}$ and the exciton densities considered here are much smaller than the Mott density $\sim a^{-2}$, the correction to the splitting is supposed to be much smaller than its value without the renormalization.

Futhermore, we show (see Appendices~\ref{appB} and~\ref{appD}) that in the rigid $1s$-exciton limit the pure exciton-exciton contribution to interaction for Wannier-Mott excitons should be larger than that from saturation, since
the Rabi-coupling term in the saddle-point Eqs.~(\ref{saddle-point-eqA},\ref{saddle-point-eqB}) in this limit is treated as perturbation. As mentioned in the Introduction, the exciton interaction constant in this limit at $T=0$ for excitons in QWs (i.e. with Coulomb interaction potential) is equal $g_{\rm ex}= \vor{(1 - 315\pi^2/4096)4\pi\hbar^2/m} =6.06 E_{\rm b} a^2$~\cite{tokatly, yamamoto,schwendimann,binder_scat, combescot,glazov,parish_ehph1}.
For TMD-based materials (with Rytova-Keldysh interaction potential) at $T=0$ we derive from Eq.~(\ref{gex}) $g_{\rm ex} = (16/\pi) e^2 a f(r_0/a)$, where the overlap $f(r_0/a)$ of the $1s$--exciton wavefunction and Fourier image of the interaction potential depends only on the ratio between the screening length $r_0$ and the Bohr radius $a$. Importantly, due to the reduced screening and hence small spread of the wavefunction, this overlap is small and decreasing with the growth of $r_0$. As a consequence, the exciton interaction constant $g_{\rm ex}$ for 2D materials with the Rytova-Keldysh interaction between the charges is smaller than that for the 2D Coulomb potential.

It is important to underline that, as can be seen from Eqs.~(\ref{gex}) and (\ref{gsat}) even without the substitution of $\chi({\bf p})$ in the hydrogenic shape, the constants $g_{\rm ex}$ and $g_{\rm sat}$ are defined solely by the properties of the material and cannot be used as tunable fitting parameters, unless one assumes that the rigid exciton limit is violated and thus the theoretical expressions obtained in this limit are not valid. In this light, the conclusions drawn in Ref.~\cite{menon} from direct measurements of the Rabi splitting at different exciton densities in monolayer WS$_2$ that saturation in their system is huge, requires reconsideration. Furthermore, the recent experiment~\cite{stepanov} performed for monolayer MoSe$_2$ revealed from fitting $g_{\rm sat}\gg g_{\rm ex}$ (in~\cite{menon}, $g_{\rm ex}$ is totally disregarded). However, since the excitons in TMD materials are robust ($E_{\rm b}\gg\hbar \Omega_{\rm R}$), the saturation process should be a correction rather than the leading term. We note that for the case of GaAs QWs~\cite{richard} where $E_{\rm b} \gtrsim \hbar\Omega_{\rm R}$, the exciton rigidity can be undermined by the presence of photons and $g_{\rm sat}$ could be of the same order with $g_{\rm ex}$, but the effect of flexibility (mixing of the excitonic $s$-series) in TMDs is highly unlikely.

\begin{table}[t]
\centering
\begin{tabular}{|P{.8 cm} P{1.2 cm}|P{2 cm}|P{2 cm}|P{1.5 cm}|}
\hline
\multicolumn{2}{|l|}{ \quad Material  }  &  $g_{\rm ex}$, $\mu$eV$\mu$m$^2$   &  $g_{\rm sat}$,  $\mu$eV$\mu$m$^2$ & $T$  \\ \hline
\multicolumn{1}{|l|}{\multirow{4}{*}{ WS$_{2}$}} & exp. \cite{menon}  &  -- & $10.0\pm 0.4$  & room\\
\multicolumn{1}{|l|}{}               &  theor. & $1.87 $ & $0.17$  &-- \\ \cline{2-5}
\multicolumn{1}{|l|}{}  & exp. \cite{zhao} &  $ 0.055\pm 0.015~$ & $0.11\pm 0.035$  & room  \\
 \multicolumn{1}{|l|}{}           &  theor. & $1.87 $ & $0.2$ & --
\\ \hline
\multicolumn{1}{|l|}{\multirow{4}{*}{MoSe$_2$}} & exp.  \cite{stepanov}  &$ 4.3 \pm 4$& $3.2\pm 0.8$ & room  \\
\multicolumn{1}{|l|}{}                  & theor. &  $0.96 $&  $0.06$ & --\\   \cline{2-5}
\multicolumn{1}{|l|}{} & exp. \cite{deng_const} & $0.12\pm 0.01$ &    $0.015 \pm 0.003$ & 5~K  \\
\multicolumn{1}{|l|}{}   & theor. &  $0.76 $&  $0.04$ & --\\ \hline
\end{tabular}
\caption{Comparison of experimental and theoretical values for $g_{\rm ex}$ and $g_{\rm sat}$ in TMD monolayers, with the theoretical values obtained in the current work.}
\label{table_const}
\end{table}
In Table~\ref{table_const}, we summarize the existing experimental estimates for $g_{\rm ex}$ and $g_{\rm sat}$ obtained from fitting of the polariton branches in TMD monolayers, and compare them to each other and to the theoretical values obtained here (for interaction constants obtained from numerical wavefunctions see \cite{numericalWF}). One sees that only the experimental values obtained at low temperature (5~K) in Ref~\cite{deng_const} have the same order as the theoretically-predicted constants. Hence, noting that straightforward introduction of the interaction constants works only at $T\to 0$ when one can factorize the exciton field (since $\chi_{\bf k}({\bf p})$ stops being dependent on $\Omega$), one needs to rely on the full temperature-dependent treatment [see Eq.~(\ref{D0})]. More strikingly, the nonlinearities experimentally reported for the same materials in different works (even at the same $T$) differ by orders of magnitude. We conclude that while the rigid-exciton limit for TMD-materials is expected to be valid with a great accuracy, there are other mechanisms at play strongly altering the polariton nonlinearities.

\section{Renormalization of the Rabi splitting due to dark excitons} \label{sec_rabi}
\vspace{-10pt}
In this section, we show that one of such possible mechanisms to explain this inconsistency can be addressed by taking into account the spins of particles and considering the influence of dark states on the saturation process. For simplicity, we restrict ourself only to the case of zero temperature. The details of derivations, which in spirit are analogous to those described in Sec.~\ref{sec_action}, are presented in Appendix~\ref{appD}. In this case, four excitonic fields are introduced, two of which correspond to bright $\Delta_{\pm 1}$ and the other two to dark excitons $\Delta_{\rm d,\bar{d}}$ (here ${\rm d}$, $\bar{\rm d}$ generically denote the two spin projections of excitons that are not coupling to light, i.e. $\pm2$ for GaAs and $0$ for TMD excitons in two different valleys). Moreover, in Appendix~\ref{appD} we take into account possible spin-bands splitting, i.e. different dispersion curvatures for electrons with different spin projections. The resulting saddle-point equations for the four excitonic fields are differ from each other due to the deviation in the exciton dispersions and, consequently, in the $1s$--exciton wavefunctions. As a result, the exciton interaction and saturation constants for $\Delta_{\pm1}$ and $\Delta_{\rm d,\bar{d}}$ generally differ as well.

Here, to show the simplest case, we assume that photons of only one polarization (either $+1$ or $-1$) are present in the system and that the effective electron masses $m_{c(v)}$ corresponding to different spin-bands are equal. Then one obtains from the general expression the splitting between the UP and LP branches at ${{\bf k}=0}$:
\begin{multline}\label{rabi_dark}
    \hbar\Omega_{R}(n_{\rm ex}, n_{\rm d}) = \hbar\Omega_{\rm R}\left[1  -  \frac{g_{\rm sat}(4 n_{\rm ex} + 2 n_{\rm d})}{\hbar \Omega_{R}} \right.\\ + \left.\frac{g^2_{\rm sat}(n_{\rm ex} \!+ n_{\rm d})(3n_{\rm ex} \!+ n_{\rm d}) \!+\! g^2_{\rm ex}(n_{\rm ex} \!+ n_{\rm d})^2}{(\hbar \Omega_{\rm R})^2}\right]^{\!1/2}\!\!\!\!,
\end{multline}
where $n_{\rm d} = \overline{C}_{\bar{\rm d}} C_{\bar{\rm d}}+ \overline{C}_{\rm d} C_{\rm d}$ is the dark exciton density, $C_{\rm d(\bar{d})}$ is the dark-exciton field depending only on the total exciton momentum. Strictly speaking, in Eq.~(\ref{rabi_dark}) we took into account only the dominant terms, \vor{yet it is important to underline that the obtained splitting accounts for both the interactions that blueshift both the LP and UP branches and the saturation which results in the blueshift of LP and the redshift of UP} (for details see Appendix \ref{appD}). Surprisingly, $n_{\rm ex}$ and $n_{\rm d}$ enter this expression additively with the coefficients of the same order, which underlines that the dark exciton population cannot be disregarded when addressing the exciton saturation. The presence of dark exciton states, as clearly seen, leads to further quench of the Rabi splitting with the growth of both densities, which could be the possible explanation of the abovementioned experimental observations. It is worth noting that a similar model including the reservoir contribution was  phenomenologically proposed in Ref.~\cite{opala}, while the reservoir was assumed to result from biexciton states.
Our model, however, does not take the exciton bound states into account, thus overlooking the interactions of excitons with antiparallel spins $g_{\rm ex}^{\uparrow\downarrow} |C_{+1}|^2|C_{-1}|^2$ which can contribute to polariton interactions~\cite{glazov,stepanov}. In the existing literature~\cite{vladimirova,parish_pol,wouters_biex,parish_biex} the corresponding interaction constant $g_{\rm ex}^{\uparrow\downarrow}$ is demonstrated to be negative and, in general, comparable with $g_{\rm ex}$. Nevertheless, on the qualitative level, these attractive interactions are expected to cause the decrease of blue- and redshifts of the polariton dispersions. 

\section{Conclusions} \label{sec_concl}
\vspace{-10pt}
To summarize, we developed the approach to perturbatively describe the electron-hole-photon system in presence of strong coupling, based on the equilibrium path integral technique. This approach, in general, \vor{is a first-time description of exciton-polariton nonlinearities that} allows to track different channels of dual-field pairing and self-consistently \vor{account for} finite temperatures and the full exciton $\nu s$-series. It can be applied to any material system (such as conventional quantum-well microcavities or those based on TMD monolayers and bilayers). The difference in the charges interaction potentials will enter via the shape of the exciton wavefunctions that define the final quantitative results. As a logical extension we also provided (see Appendix~\ref{appD}) the treatment of particle spins, which revealed the contributions to exciton nonlinearities due to bright-bright, bright-dark, dark-dark exciton interactions and spin-flip processes.

In the simplified case of rigid excitons, our analysis provides \vor{several} important results relevant to understanding the giant nonlinearities observed in TMD-based microcavities. First, we show that finite temperatures result in a sizeable change of the interaction constants and the usual zero-$T$ expressions cannot be used for the estimates in room-temperature studies. Second, we reveal that the exciton-photon conversion is affected by the presence of the reservoir particles (such as dark excitons). We rigorously derive the terms involving the dark exciton fields in effective action, and the dark-exciton density contributions to the Rabi splitting renormalization. These results call for a detailed  comparative experimental studies of the exciton saturation in TMD structures, to be performed at different temperatures, which would allow to elucidate the relative influence of the two factors on the observed nonlinearity.

In a more sophisticated case of flexible excitons, which is more relevant to quantum-well polaritons or for \vor{systems with} very strong coupling, an important conclusion is that the concept of interaction constants describing the two sources of nonlinearity in the system starts to fail. As several exciton $s$-states start to contribute to the exciton field, the relative motion of the electron and hole within the exciton cannot be separated from the motion of the exciton as a whole. The nonlinearities arising in this case from the fourth-order term with respect to the exciton field are not factorized to the integral over the exciton fields and wavefunctions overlap integrals (which yield the above mentioned interaction and saturation constants). Furthermore, the Bethe-Salpeter equation in this case\vor{, while having formally the same shape as known from the semiconductor literature,} contains not only the temperature dependence, but also the contribution of the momentum-dependent photon energy dispersion to the formation of the exciton.

We note that our study did not consider additional contributions to the photon-exciton action such as the formation of multi-exciton bound states. These scenarios were addressed in recent works~\cite{parish_biex,shelykh_biex} where it was shown that biexciton-polariton formation is possible, and that biexciton transition therefore can also alter nonlinear optical response of TMDs.

\section*{Acknowledgements}
\vspace{-10pt}
The authors are thankful to Vanik Shahnazaryan for useful discussions. The partial financial support from the NRNU MEPhI Priority 2030 Program is acknowledged. The work of A.G. is funded by the Foundation for the Advancement of Theoretical Physics and Mathematics ``BASIS'', under the Grant No.~22--1--5--30--1.

\appendix
\begin{widetext}
\section{Additional fields (other channels of pairing)}\label{appA}
In the beginning of the main text we introduce the exciton field only. Strictly speaking, 
such consideration overlooks the exciton interaction mediated by electron scattering processes. In this Appendix, we address this problem and show how taking into account the 
off-diagonal density field $\overline{\Psi}_{i}({\bf r}, \tau)\Psi_{i}({\bf r}^\prime\!, \tau)$ ($i=c,v$) changes the exciton interaction term. For clarity, in the derivations of this section we will omit the ``photon'' part of Eq.~(\ref{action_e-h-ph_initial}) which does not play a role here, being not coupled with the density fields.
Turning to the initial form of the action (\ref{action_e-h-ph_initial}), we introduce not only the previously discussed fields $\Delta$ and $\phi$, but also other density-like fields which arise when considering other pairing channels, such as the off-diagonal $\Phi_i({\bf r}, {\bf r}^\prime\!, \tau) = \overline{\Psi}_{i}({\bf r}, \tau)\Psi_{i}({\bf r}^\prime\!, \tau)$ and diagonal $\xi_{i}({\bf r}, \tau) = \overline{\Psi}_{i}({\bf r}, \tau)\Psi_{i}({\bf r}, \tau)$ density fields, using the Hubbard-Stratonovich transformation~\cite{HS}. 

First, we define the momentum-frequency representation for the considered fields:
\begin{equation}
    \Phi_{i}(x, y) = \frac{\sqrt{T}}{S}\sum_{{\bf k},{\bf k}^\prime} \sum_{\,\,\Omega_{n}} \Phi_i({\bf k}, {\bf k^\prime}\!, \Omega_n) e^{i{\bf k}\cdot{\bf x} - i {\bf k^\prime}\!\!\cdot{\bf y} - i\Omega_{n} \tau}
\end{equation}
\begin{equation}
    \xi_{i}(x) = \sqrt{\frac{T}{S}}\sum_{{\bf k},{\bf k}^\prime} \sum_{\,\,\Omega_{n}} \xi_i({\bf k}, {\bf k^\prime}\!, \Omega_n) e^{i{\bf k}\cdot{\bf x} - i{\bf k^\prime}\!\!\cdot{\bf x}- i\Omega_{n} \tau},
\end{equation}
where ${\bf k}$ are discrete momenta and $\Omega_{n}  = 2\pi n T \equiv \Omega$ are bosonic Matsubara frequencies. As one can see, the Fourier image of the diagonal density field depends on one momentum (${\bf k}-{\bf k}^\prime$) only, which leads, as we will see further, to qualitatively different contributions to the exciton interaction.
In the frequency-momentum representation the action takes the form:
\begin{multline}
    \mathcal{S}[\Psi_{\!c},\Psi_{\!v}, \phi, \Delta, \Phi_{c}, \Phi_{v}, \xi_c, \xi_v]=\sum_{{\bf k},\omega}\left[
       \begin{pmatrix}
            \overline{\Psi}_{\! c}(k) & \overline{\Psi}_{\! v}(k)
        \end{pmatrix} \!\!
        \begin{pmatrix}
            -i \omega + \varepsilon_c({\bf k}) - \mu_{c} & 0 \\ 0 &   -i \omega + \varepsilon_v({\bf k}) - \mu_{v}
        \end{pmatrix} \!\!
        \begin{pmatrix}
             {\Psi}_{\! c}(k) \\ {\Psi}_{\! v}(k)
        \end{pmatrix}
    \right] \\[4pt]
    -i \! \sum_{\substack{{\bf k}_1,{\bf k}_2\\\Omega}} \!\! \left[ \bar{\phi}({\bf k}_1,{\bf k}_2, \Omega)\Delta({\bf k}_1,{\bf k}_2, \Omega) + \bar{\Delta}({\bf k}_1,{\bf k}_2, \Omega) \phi({\bf k}_1,{\bf k}_2, \Omega) \right] - \!\!\! \sum_{\substack{{\bf k}_1\dots{\bf k}_4\\ \Omega}} \!\!\!\! V({\bf k}_1 - {\bf k}_2) \bar{\Delta}({\bf k}_1,{\bf k}_4,\Omega) \Delta({\bf k}_2,{\bf k}_3,\Omega) \delta({\bf k}_1+{\bf k}_3, {\bf k}_2+{\bf k}_4)
    \\
    + \frac{1}{2}\sum_{{\bf k}, \Omega} 
    V^{-1}({\bf k})\xi({\bf k}, \Omega)\xi(-{\bf k}, -\Omega) + \frac{1}{2} \! \sum_{\substack{{\bf k}_1\dots{\bf k}_4\\ \Omega}} \!  \sum\limits_{\,\,i=c,v} V^{-1}({\bf k}_1 - {\bf k}_2) \Phi_i({\bf k}_1,{\bf k}_4,\Omega) \Phi_i({\bf k}_3,{\bf k}_2,\Omega) \delta({\bf k}_1+{\bf k}_3, {\bf k}_2+{\bf k}_4) \\
    + \sqrt{T} \sum_{{\bf k}_1,\omega_{1}}\sum_{{\bf k}_2,\omega_2} \!\! \left[
        \begin{pmatrix}
            \overline{\Psi}_{\! c}(k_1) & \overline{\Psi}_{\! v}(k_1)
        \end{pmatrix} \!\!
        \begin{pmatrix}
           \Phi_{c}(k_2, k_1) + i\xi(k_2- k_1)&i \phi({\bf k}_1,{\bf k}_2, \omega_{1} - \omega_{2}) \\
          i \bar{\phi}({\bf k}_2,{\bf k}_1, \omega_{2} - \omega_{1}) & \Phi_{v}(k_2,k_1) +  i\xi(k_2- k_1)
        \end{pmatrix} \!\!
        \begin{pmatrix}
             {\Psi}_{\! c}(k_2) \\ {\Psi}_{\! v}(k_2)
        \end{pmatrix}\right] + \dots,
\end{multline}
where, compared to Eq.~(\ref{action1}) the terms containing $\Psi_{\rm ph}$ are omitted ($\dots$) and the notation $\xi = \xi_{c} + \xi_{v}$ is introduced.
Although the expression is quite cumbersome, all the contributions containing $\Phi$ and $\xi$ will be integrated out, and the final expression remains elegant.
After performing the path integration over the fermionic fields, the second order of the ${\rm Tr ln}\mathcal{G}^{-1}$ expansion series yields the relation between $\phi$ and $\Delta$, as in the main text,
\begin{equation*}
    \phi({\bf k}_1,{\bf k}_2,\Omega) = \frac{-i}{\mathcal{A}_{{\bf k}_1{\bf k}_2}^{\,\,\Omega}}\Delta({\bf k}_1,{\bf k}_2,\Omega),
\end{equation*}
and renormalizes the Coulomb interaction for the 
off-diagonal and diagonal densities
\begin{eqnarray}
    V^{-1}({\bf k}_1 \!-\! {\bf k}_2) &\to& V^{-1}({\bf k}_1 \!-\! {\bf k}_2) + T \sum_{\omega^\prime} \! \mathcal{G}_{0}({\bf k}_1, \omega^\prime \!+ \Omega)_{11} \mathcal{G}_{0}({\bf k}_2, \omega^\prime)_{11}\quad \text{for }\Phi_c, \nonumber \\
    V^{-1}({\bf k}_1 \!-\! {\bf k}_2) &\to& V^{-1}({\bf k}_1 \!-\! {\bf k}_2) + T \sum_{\omega^\prime} \! \mathcal{G}_{0}({\bf k}_1, \omega^\prime \!+ \Omega)_{22} \mathcal{G}_{0}({\bf k}_2, \omega^\prime)_{22}\quad \text{for }\Phi_v, \label{renorm} \\
    V^{-1}({\bf k}_1 \!-\! {\bf k}_2) &\to& V^{-1}({\bf k}_1 \!-\! {\bf k}_2) \!-\! T \! \sum_{\omega^\prime} \! \mathcal{G}_{0}({\bf k}_1, \omega^\prime \!+ \Omega)_{11} \mathcal{G}_{0}({\bf k}_2, \omega^\prime)_{11} \!+\! T\! \sum_{\omega^\prime} \! \mathcal{G}_{0}({\bf k}_1, \omega^\prime \!+ \Omega)_{22} \mathcal{G}_{0}({\bf k}_2, \omega^\prime)_{22}\quad \text{for }\xi, \nonumber
\end{eqnarray}
where $\omega^\prime = (2n+1)\pi/\beta$ is the fermionic Matsubara frequency.

The renormalization (\ref{renorm}) is assumed to be small, since we assume the small-parameter expansion. We note that since the variation of the Green's function $\delta \mathcal{G}^{-1}$ contains both diagonal and off-diagonal elements, the first- and third-order terms appear in the expansion series. The first-order term is zero under the assumption that the system is electrically neutral.
The third-order term has the form:
\begin{multline}
    \frac{1}{3}{\rm Tr}(\mathcal{G}_0\delta\mathcal{G}^{-1})^3 = \frac{1}{3} \!\! \sum_{k_{1}, k_{2}, k_{3}} \!\! \mathcal{G}_{0}(k_1)_{\alpha\alpha}\delta\mathcal{G}^{-1}(k_1,k_2)_{\alpha \beta}\mathcal{G}_0(k_2)_{\beta \beta}\delta\mathcal{G}^{-1}(k_2, k_3)_{\beta \gamma}\mathcal{G}_0(k_3)_{\gamma \gamma}\delta\mathcal{G}^{-1}(k_3, k_1)_{\gamma \alpha} \\
    = \sqrt{T} \!\! \sum_{{\bf k}_{1},{\bf k}_{2},{\bf k}_{3}} \! \sum_{\Omega_1, \Omega_2}   \! \bigl[\tilde{\Pi}_{121}\bigr]_{{\bf k}_{1}{\bf k}_{2}{\bf k}_{3}}^{\,\Omega_1 \Omega_2}\, \Delta({\bf k}_1,{\bf k}_2, \Omega_1) \bar{\Delta}({\bf k}_3, {\bf k}_2, \Omega_2) \!\left[\Phi_c({\bf k}_1, {\bf k}_3, \Omega_1 - \Omega_2)+ i\xi({\bf k}_1- {\bf k}_3, \Omega_1 - \Omega_2)\right] \\
    + \sqrt{T} \!\! \sum_{{\bf k}_{1},{\bf k}_{2},{\bf k}_{3}} \sum_{\Omega_1, \Omega_2} \! \bigl[\tilde{\Pi}_{122}\bigr]_{{\bf k}_{1}{\bf k}_{2}{\bf k}_{3}}^{\,\Omega_1 \Omega_2}\,\Delta({\bf k}_1,{\bf k}_2, \Omega_1) \bar{\Delta}({\bf k}_1, {\bf k}_3, \Omega_2) \!\left[\Phi_v({\bf k}_3, {\bf k}_2, \Omega_1 - \Omega_2) + i \xi({\bf k}_3- {\bf k}_2, \Omega_1 - \Omega_2)\right],
\end{multline}
where $\alpha,\beta,\gamma$ run over the values $\overline{1,2}$, but we choose only those that contain the fields $\phi$ (or $\Delta$), hence the notation for the third-order loops $\Pi_{121}$ and $\Pi_{122}$ is introduced (see Appendix~\ref{appE} for definitions and details of calculation). Tilde, as before, denotes the correspondence of a loop to the exciton field $\Delta$ (i.e. after the exclusion of the auxiliary field $\phi$).

Integrating over the fields $\Phi_{c,v}$ we obtain the correction to the exciton interaction due to the electron-electron scattering (i.e. screening):
\begin{multline}\label{S'scr}
    \mathcal{S}^\prime_{\rm scr} = - \frac{T}{2} \sum_{{\bf q},{\bf k}_1\dots{\bf k}_4} \sum_{\substack{\Omega_1,\Omega_2 \\ \Omega_1^\prime,\Omega_2^\prime}} \! V({\bf q}) \bigl[\tilde{\Pi}_{121}\bigr]_{{\bf k}_1 {\bf k}_3{\bf k}_2}^{\,\Omega_1\Omega_2} \bigl[\tilde{\Pi}_{121}\bigr]_{{\bf k}_2 - {\bf q}, {\bf k}_4, {\bf k}_1 - {\bf q}}^{\,\,\Omega_1^\prime\,\,\Omega_2^\prime}\,\Delta({\bf k}_1,{\bf k}_3, \Omega_1)\bar{\Delta}({\bf k}_2,{\bf k}_3, \Omega_2)\Delta({\bf k}_2 - {\bf q}, {\bf k}_4, \Omega_1^\prime) \\[-5pt]
    \times  \bar{\Delta}({\bf k}_1-{\bf q}, {\bf k}_4, \Omega_2^\prime)\delta_{\Omega_1 + \Omega_1^\prime ,\Omega_2 + \Omega_2^\prime} 
    -\frac{T}{2} \!\! \sum_{{\bf q},{\bf k}_1\dots{\bf k}_4} \sum_{\substack{\Omega_1, \Omega_2 \\
    \Omega_1^\prime, \Omega_2^\prime}} \! V({\bf q}) \bigl[\tilde{\Pi}_{122}\bigr]_{{\bf k}_3 {\bf k}_2{\bf k}_1}^{\, \Omega_1 \Omega_2} \bigl[\tilde{\Pi}_{122} \bigr]_{ {\bf k}_4, {\bf k}_1 - {\bf q}, {\bf k}_2 - {\bf q}}^{\,\,\Omega_1^\prime\,\, \Omega_2^\prime} \Delta({\bf k}_3,{\bf k}_2, \Omega_1) \\ 
   \times \bar{\Delta}({\bf k}_3,{\bf k}_1, \Omega_2)\Delta({\bf k}_4, {\bf k}_1 - {\bf q}, \Omega_1^\prime)\bar{\Delta}({\bf k}_4, {\bf k}_2 - {\bf q},\Omega_2^\prime) \delta_{\Omega_1 + \Omega_1^\prime ,\Omega_2 + \Omega_2^\prime}.
\end{multline}
Eq.~(\ref{S'scr}) has the same structure as the fourth-order term in the ${\rm Trln}\mathcal{G}^{-1}$ expansion [see Eq.~(\ref{phi4})] and therefore is a correction to the $|\Delta|^4$-term.
In a similar fashion, integrating out the diagonal density fields $\xi_{c,v}$ leads to the $|\Delta|^4$-term describing the repulsive exciton-exciton interaction
\begin{multline}\label{S'vdW}
   \mathcal{S}^\prime_{\rm vdW} = \frac{T}{2} \!\!\sum_{{\bf q},{\bf k}_1\dots{\bf k}_4} \sum_{\substack{\Omega_1, \Omega_2 \\ \Omega_1^\prime,\Omega_2^\prime}} \! V({\bf q}) \bigl[\tilde{\Pi}_{121}\bigr]_{{\bf k}_1, {\bf k}_2, {\bf k}_1 - {\bf q}}^{\,\, \Omega_1\,\, \Omega_2} \bigl[\tilde{\Pi}_{121}\bigr]_{{\bf k}_3, {\bf k}_4, {\bf k}_3 + {\bf q}}^{\,\, \Omega_1^\prime \,\, \Omega_2^\prime}\, \Delta({\bf k}_1,{\bf k}_2, \Omega_1) \bar{\Delta}({\bf k}_1 - {\bf q},{\bf k}_2 , \Omega_2)\Delta({\bf k}_3,{\bf k}_4, \Omega_1^\prime) \\[-5pt]
   \times \bar{\Delta}({\bf k}_3 + {\bf q},{\bf k}_4, \Omega_2^\prime)\delta_{\Omega_1 + \Omega_1^\prime, \Omega_2 + \Omega_2^\prime}
   +\frac{T}{2} \!\! \sum_{{\bf q}, {\bf k}_1\dots{\bf k}_4} \sum_{\substack{\Omega_1, \Omega_2 \\ \Omega_1^\prime,\Omega_2^\prime}} \! V({\bf q}) \bigl[\tilde{\Pi}_{122}\bigr]_{{\bf k}_1, {\bf k}_2, {\bf k}_2-{\bf q}}^{\,\, \Omega_1 \,\, \Omega_2} \bigl[\tilde{\Pi}_{122}\bigr]_{{\bf k}_3, {\bf k}_4, {\bf k}_4 + {\bf q}}^{ \,\, \Omega_1^\prime \,\, \Omega_2^\prime}  \Delta({\bf k}_1,{\bf k}_2, \Omega_1)
   \\
   \times \bar{\Delta}({\bf k}_1,{\bf k}_2 - {\bf q}, \Omega_2)\Delta({\bf k}_3,{\bf k}_4, \Omega_1^\prime) \bar{\Delta}({\bf k}_3,{\bf k}_4  + {\bf q}, \Omega_2^\prime) \delta_{\Omega_1 + \Omega_1^\prime,\Omega_2 + \Omega_2^\prime}
\end{multline}
which can be interpreted as a correction to the Coulomb-like interaction $(-V\bar{\Delta}\Delta)$, since 
it contains the direct Coulomb interaction of excitons as a whole. 
Therefore Eq.~(\ref{S'vdW}) may be treated as a correction to the generalized Bethe-Salpeter equation due to the van-der-Waals interaction
$$ \frac{\delta S}{\delta\bar{\Delta}}= \frac{1}{\mathcal{A}_{{\bf p}{\bf k}}^{\,\Omega}}\,\Delta({\bf p},{\bf k}, \Omega) - \sum_{{\bf q}}\left[V({\bf p} -{\bf q}) + \left(\frac{g_{\rm R}}{\sqrt{S}}\right)^{\!\!2}\!\frac{1}{E_{\rm ph}({\bf k}) -\mu- i \Omega} \right]\Delta( {\bf q},{\bf k},\Omega) + \frac{\delta (\mathcal{S}^\prime_{\rm vdW})}{\delta\bar{\Delta} }=0$$
which changes Eq.~(\ref{BSE}) just slightly and can be neglected. We conclude that the 
consideration of additional pairing channels in the initial action~(\ref{action_e-h-ph_initial}), which is performed here generally, leads to corrections due to the screening of the exciton-exciton interaction only.

\section{Calculation of the loop diagrams}\label{appE}

To derive the exciton propagator [the second-order loop shown in Fig.~\ref{fig:loops}(a)], we use the standard resummation procedure, i.e. the Sommerfeld-Watson transformation:
\begin{multline}
   \mathcal{A}_{{\bf k}_1{\bf k}_2}^{\,\,\Omega} = T \!\!\!\!\!\! \sum_{\omega^\prime = (2n+1)\pi T} \!\!\!\!\!\! \mathcal{G}_{0}({\bf k}_1, \omega^\prime \!+ \Omega)_{11} \mathcal{G}_{0}({\bf k}_2, \omega^\prime)_{22} = T \sum_{\omega^\prime}\frac{1}{[-i(\omega^\prime \!+ \Omega) + \varepsilon_c({\bf k}_1) - \mu_{c}][-i\omega^\prime \!+ \varepsilon_v({\bf k}_2)- \mu_{v} ]} \\
   =T\frac{1}{2\pi i}\frac{1}{T}\oint dz \, \frac{1}{e^{\beta z}+1} \frac{1}{[-z -i \Omega + \varepsilon_c({\bf k}_1)- \mu_{c}][-z + \varepsilon_v({\bf k}_2)- \mu_{v}]} \\[5pt]
   =\frac{1}{\varepsilon_c({\bf k}_1) - \varepsilon_v({\bf k}_2) -\mu_{c} + \mu_{v}- i \Omega}\left(\frac{1}{e^{\beta[\varepsilon_v({\bf k}_2) - \mu_{v}]}+1} - \frac{1}{e^{\beta[\varepsilon_c({\bf k}_1) - \mu_{c}]}+1}\right) = \frac{ n_{v}({\bf k}_2)-n_{c}({\bf k}_1)}{\varepsilon_c({\bf k_1}) - \varepsilon_v({\bf k}_2) -\mu - i \Omega} \,,
\end{multline}
as given in Eqs.~(\ref{A7})--(\ref{Ak1k2}) of the main text.

\begin{figure}[b]
    \centering
    \includegraphics[width=\textwidth]{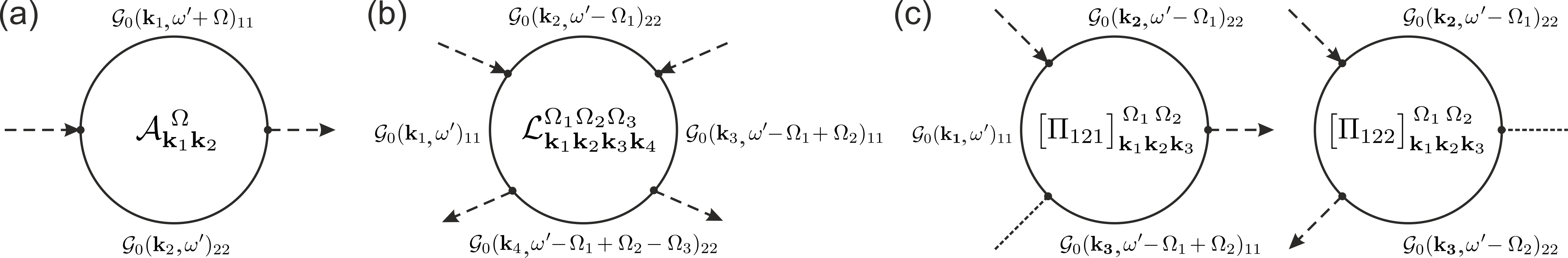}
    \caption{\small Loop diagrams corresponding to the second-order (a), fourth-order (b) and third-order (c) terms of the ${\rm Trln}\mathcal{G}^{-1}$ expansion. The solid lines represent the Green's functions as marked on the panels, the ingoing (outgoing) dashed lines correspond to the fields $\phi$ ($\bar{\phi}$), the dotted lines in (c) represent the additional fields $\Phi_{c,v}+i\xi$ (see Appendix~\ref{appA}). Summation over the fermionic Matsubara frequency $\omega^\prime$ is assumed. For the similar loops denoted with tilde (see text), the fields $\phi$ are expressed in terms of the excitonic field $\Delta$.}
    \label{fig:loops}
\end{figure}

Analogously we perform the calculation of the loop for the fourth-order term, presented in Fig.~\ref{fig:loops}(b):
\begin{multline}
    \mathcal{L}_{{\bf k}_1{\bf k}_2{\bf k}_3{\bf k}_4}^{\Omega_1\Omega_2\Omega_3} =T \sum_{\omega'}
    \mathcal{G}_{0}({\bf k}_1, \omega^\prime)_{11}\mathcal{G}_0({\bf k}_2, \omega^\prime \!- \Omega_1)_{22}\mathcal{G}_0({\bf k}_3, \omega^\prime \!- \Omega_1 + \Omega_2)_{11}\mathcal{G}_0({\bf k}_4, \omega^\prime \! - \Omega_1 + \Omega_2 - \Omega_3)_{22} \\
    = T\frac{1}{2\pi i} \frac{1}{T} \oint dz \frac{1}{e^{\beta z}+1}\frac{1}{[-z  + \varepsilon_c({\bf k}_1)- \mu_{c}][-z + i \Omega_1 + \varepsilon_v({\bf k}_2)- \mu_{v}]} \hspace{120pt} \\[5pt]
    \hspace{100pt} \times \frac{1}{[-z + i \Omega_1 - i\Omega_2 + \varepsilon_c({\bf k}_3)- \mu_{c}][-z + i \Omega_1 - i\Omega_2 + i \Omega_3 +\varepsilon_v({\bf k}_4)- \mu_{v}]}
    \\[5pt]
    =-\left\{\frac{n_{c}({\bf k}_1)}{[i\Omega_1 + \varepsilon_v({\bf k}_2)-\varepsilon_c({\bf k}_1)+ \mu][i\Omega_1- i \Omega_2 + \varepsilon_c({\bf k}_3)-\varepsilon_c({\bf k}_1)][i\Omega_1- i \Omega_2 +i \Omega_3 + \varepsilon_v({\bf k}_4)-\varepsilon_c({\bf k}_1) +\mu]} \right.\\[5pt]
    +\frac{n_{v}({\bf k}_2)}{[-i\Omega_1 + \varepsilon_c({\bf k}_1)-\varepsilon_v({\bf k}_2) - \mu][- i \Omega_2 + \varepsilon_c({\bf k}_3)-\varepsilon_v({\bf k}_2)-\mu][- i \Omega_2 +i \Omega_3 + \varepsilon_v({\bf k}_4)-\varepsilon_v({\bf k}_2)]}\\[5pt]
    + \frac{n_{c}({\bf k}_3)}{[-i\Omega_1 + i\Omega_2+ \varepsilon_c({\bf k}_1)-\varepsilon_c({\bf k}_3)][i \Omega_2 + \varepsilon_v({\bf k}_2)-\varepsilon_c({\bf k}_3)+ \mu][i \Omega_3 + \varepsilon_v({\bf k}_4)-\varepsilon_c({\bf k}_3)+\mu]}\\[5pt]
    \left.+\frac{n_{v}({\bf k}_4)}{[-i\Omega_1 + i\Omega_2- i\Omega_3+ \varepsilon_c({\bf k}_1)-\varepsilon_v({\bf k}_4)-\mu][i \Omega_2 -i\Omega_3+ \varepsilon_v({\bf k}_2)-\varepsilon_v({\bf k}_4)][-i \Omega_3 + \varepsilon_c({\bf k}_3)-\varepsilon_v({\bf k}_4)-\mu]}
    \right\} \\[5pt]
    =-\frac{1}{i\Omega_1- i \Omega_2 + \varepsilon_c({\bf k}_3)-\varepsilon_c({\bf k}_1)}\left\{\frac{n_{c}({\bf k}_1)}{[i\Omega_1 + \varepsilon_v({\bf k}_2)-\varepsilon_c({\bf k}_1)+\mu][i\Omega_1- i \Omega_2 +i \Omega_3 + \varepsilon_v({\bf k}_4)-\varepsilon_c({\bf k}_1)+\mu]}\right.
    \\[5pt]
   \hspace{180pt} \left.-\frac{n_{c}({\bf k}_3)}{[i \Omega_2 + \varepsilon_v({\bf k}_2)-\varepsilon_c({\bf k}_3)+\mu][i \Omega_3 + \varepsilon_v({\bf k}_4)-\varepsilon_c({\bf k}_3)+\mu]}\right\}
    \\[5pt]
    - \frac{1}{- i \Omega_2 +i \Omega_3 + \varepsilon_v({\bf k}_4)-\varepsilon_v({\bf k}_2)} \left\{\frac{n_{v}({\bf k}_2)}{[-i\Omega_1 + \varepsilon_c({\bf k}_1)-\varepsilon_v({\bf k}_2)-\mu][- i \Omega_2 + \varepsilon_c({\bf k}_3)-\varepsilon_v({\bf k}_2)-\mu]} \right. \hspace{20pt} \\[5pt]
    -\left. \frac{n_{v}({\bf k}_4)}{[-i\Omega_1 + i\Omega_2- i\Omega_3+ \varepsilon_c({\bf k}_1)-\varepsilon_v({\bf k}_4)-\mu][-i \Omega_3 + \varepsilon_c({\bf k}_3)-\varepsilon_v({\bf k}_4)-\mu]}\right\}.
\end{multline}
Introducing for convenience the notation
\begin{eqnarray*}
    B_{k_1 k_2} &= -i\Omega_1 + \varepsilon_{c}({\bf k}_1) - \varepsilon_{v}({\bf k}_2)-\mu, \; &
    B_{k_3 k_2} = -i\Omega_2 + \varepsilon_{c}({\bf k}_3) - \varepsilon_{v}({\bf k}_2) -\mu,\\
    B_{k_3 k_4} &= -i\Omega_3 + \varepsilon_{c}({\bf k}_3) - \varepsilon_{v}({\bf k}_4) -\mu, \;
    &B_{k_1 k_4} = -i\Omega_1 + -i\Omega_2 -i\Omega_3 + \varepsilon_{c}({\bf k}_1) - \varepsilon_{v}({\bf k}_4)-\mu,
\end{eqnarray*}
we express the fourth-order loop in the following form:
\begin{equation}
   \mathcal{L}_{{\bf k}_1{\bf k}_2{\bf k}_3{\bf k}_4}^{\Omega_1\Omega_2\Omega_3} = \frac{1}{B_{k_1k_2} \!-\! B_{k_3 k_2}} \! \left[\frac{n_{c}({\bf k}_1)}{B_{k_1k_2} B_{k_1k_4}} - \frac{n_{c}({\bf k}_3)}{B_{k_3k_2}B_{k_3k_4}}\right] \!-\! \frac{1}{B_{k_3k_2} - B_{k_3k_4}}\!\left[\frac{n_{v}({\bf k}_2)}{B_{k_1k_2}B_{k_3k_2}} - \frac{n_{v}({\bf k}_4)}{B_{k_1k_4}B_{k_3k_4}}\right].
\end{equation}

The third-order loops that are of interest to our calculations presented in Appendix~\ref{appA} are shown in Fig.~\ref{fig:loops}(c) and are defined as
\begin{eqnarray}
     \bigl[\Pi_{121}\bigr]_{{\bf k}_1{\bf k}_2{\bf k}_3}^{\,\Omega_1 \, \Omega_2} &=& T\sum_{\omega^\prime}\mathcal{G}_{0}({\bf k_1}, \omega^\prime)_{11} \mathcal{G}_0({\bf k_2},\omega^\prime \!-\Omega_1)_{22} \mathcal{G}_0({\bf k_3}, \omega^\prime \!- \Omega_1 + \Omega_2)_{11},  \\
     \bigl[\Pi_{122}\bigr]_{{\bf k}_1{\bf k}_2{\bf k}_3}^{\,\Omega_1 \, \Omega_2} &=& T\sum_{\omega^\prime}\mathcal{G}_{0}({\bf k_1}, \omega^\prime)_{11} \mathcal{G}_0({\bf k_2}, \omega^\prime \!- \Omega_1)_{22} \mathcal{G}_0({\bf k_3}, \omega^\prime \!- \Omega_2)_{22}.
\end{eqnarray}
The calculation is performed in the same way:
\begin{equation}
     \bigl[\Pi_{121}\bigr]_{{\bf k}_1{\bf k}_2{\bf k}_3}^{\,\Omega_1 \, \Omega_2} = T\frac{1}{2 \pi i} \frac{1}{T} \oint  dz \frac{1}{e^{\beta z}+1} \,\frac{1}{-z \!+\! \varepsilon_{c}({\bf k}_1) \!-\! \mu_{c}} \,\frac{1}{-z \!+\! i \Omega_1 \!+\! \varepsilon_{v}({\bf k}_2) \!-\! \mu_{v}} \, \frac{1}{-z \!+\! i \Omega_1 \!-\! i \Omega_2  \!+\! \varepsilon_{c}({\bf k}_3) \!-\! \mu_{c}}
     \nonumber 
\end{equation}
\begin{multline}
     = - \left\{\frac{n_{c}({\bf k}_1)}{[i \Omega_1 + \varepsilon_{v}({\bf k}_2) - \varepsilon_{c}({\bf k}_1) + \mu][i \Omega_1 - i \Omega_2 +\varepsilon_{c}({\bf k}_3) - \varepsilon_{c}({\bf k}_1)]} +  \right. \hspace{93pt}
     \\[5pt]
     \left. \frac{n_{v}({\bf k}_2)}{[-i\Omega_1 \!+\! \varepsilon_{c}({\bf k}_1) \!-\! \varepsilon_{v}({\bf k}_2) \!-\! \mu][-i\Omega_2 \!+\! \varepsilon_{c}({\bf k}_3) \!-\! \varepsilon_{v}({\bf k}_2) \!-\! \mu]} + \frac{n_{c}({\bf k}_3)}{[-i \Omega_1 \!+\! i \Omega_2 \!+\! \varepsilon_{c}({\bf k}_1) \!-\! \varepsilon_{c}({\bf k}_3)][i \Omega_2 \!+\! \varepsilon_{v}({\bf k}_2) \!-\! \varepsilon_{c}({\bf k}_3) \!+\! \mu]}\right\} \\[5pt]
     = \frac{n_{c}({\bf k}_1)}{B_{k_1k_2}(B_{k_1k_2} - B_{k_3k_2})} + \frac{n_{v}({\bf k}_2)}{B_{k_1k_2} B_{k_3k_2}} + \frac{n_{c}({\bf k}_3)}{B_{k_3k_2}(B_{k_3k_2}-B_{k_1k_2})},
\end{multline}
\begin{multline}
    \bigl[\Pi_{122}\bigr]_{{\bf k}_1{\bf k}_2{\bf k}_3}^{\,\Omega_1 \, \Omega_2} = - \left\{\frac{n_{c}({\bf k}_1)}{[i \Omega_1 + \varepsilon_{v}({\bf k}_2) -\varepsilon_{c}({\bf k}_1) + \mu][i \Omega_1 - i\Omega_2 + \varepsilon_v{(\bf k}_3) - \varepsilon_{c}({\bf k}_1)+\mu]} +\right.\\[5pt]
    \left.  \frac{n_{v}({\bf k}_2)}{[-i \Omega_1 \!+\! \varepsilon_{c}({\bf k}_1) \!-\!\varepsilon_{v}({\bf k}_2) -\mu][-i \Omega_2 \!+\! \varepsilon_v{(\bf k}_3) \!-\! \varepsilon_{v}({\bf k}_2)]} + \frac{n_{v}({\bf k}_3)}{[-i \Omega_1 \!+\! i\Omega_2 \!+\! \varepsilon_{c}({\bf k}_1) \!-\! \varepsilon_{v}({\bf k}_3) -\mu][i \Omega_2 \!+\! \varepsilon_{v}({\bf k}_2) \!-\! \varepsilon_{v}({\bf k}_3)]} \right\}
    \\[5pt]
    = \frac{n_{c}({\bf k}_1)}{B_{k_1k_2}B_{k_1k_3}} +\frac{n_{v}({\bf k}_2)}{B_{k_1k_2}(B_{k_1k_2}-B_{k_1k_3})} +\frac{n_{v}({\bf k}_3)}{B_{k_1k_3}(B_{k_1k_3}-B_{k_1k_2})}.
\end{multline}

\section{Calculations at zero temperature}\label{appB}
This Appendix is devoted to simplification of the general expressions obtained in Sec.~\ref{sec_action} for the case $T=0$ which allows to obtain some analytical results comparable with those exciting in literature (see discussion in the main text).
Assuming the limit of a large system size, the Fourier transforms can be rewritten as
\begin{subequations}
    \begin{eqnarray}
    & &\Psi_{i}({\bf r}, \tau) =\int \!\! \frac{d{\bf k}}{(2\pi)^2}\frac{d\omega}{2\pi} \, \Psi_{i}({\bf k}, \omega) e^{i {\bf k}\cdot{\bf x} - i\omega\tau},\\
    & &\Delta(x, y) = \int \!\! \frac{d{\bf k}}{(2\pi)^2} \frac{d{\bf k}^\prime}{(2\pi)^2} \frac{d\Omega}{2\pi} \, \Delta({\bf k}, {\bf k}^\prime\!, \Omega) e^{i{\bf k}\cdot{\bf x} - i {\bf k}^\prime\!\cdot{\bf y} - i\Omega \tau},\\
    & &\phi(x, y) =\int \!\! \frac{d{\bf k}}{(2\pi)^2} \frac{d{\bf k}^\prime}{(2\pi)^2}\frac{d\Omega}{2\pi} \, \phi({\bf k}, {\bf k}^\prime\!, \Omega) e^{i{\bf k}\cdot{\bf x} - i {\bf k}^\prime\!\cdot{\bf y} - i\Omega \tau}.
\end{eqnarray}
\end{subequations}
At zero temperature $n_v=1$, $n_c=0$, and one can derive from the expansion series the analytical expressions for all the loops considered in Appendix~\ref{appE},
\begin{subequations}
    \begin{eqnarray}
    \mathcal{A}_{{\bf k}_1{\bf k}_{2}}^{\,\,\Omega} &=&\frac{1}{\varepsilon_c({\bf k}_1) - \varepsilon_v({\bf k}_2) - \mu- i \Omega}, \\
    \bigl[\Pi_{121}\bigr]_{{\bf k}_1{\bf k}_2{\bf k}_3}^{\,\Omega_1 \,\Omega_2} &=& \frac{1}{[-i \Omega_1 + \varepsilon_{c}({\bf k}_1) - \varepsilon_{v}({\bf k}_2) - \mu][-i \Omega_2 + \varepsilon_{c}({\bf k}_3) - \varepsilon_{v}({\bf k}_2) - \mu]}, \\
    \bigl[\Pi_{122}\bigr]_{{\bf k}_1{\bf k}_2{\bf k}_3}^{\,\Omega_1 \,\Omega_2} &=& \frac{1}{[-i \Omega_1 + \varepsilon_{c}({\bf k}_1) - \varepsilon_{v}({\bf k}_2) - \mu][-i \Omega_1 + i \Omega_2 + \varepsilon_{c}({\bf k}_1) - \varepsilon_{v}({\bf k}_3) - \mu]}.\\[4pt]
    \mathcal{L}_{{\bf k}_1{\bf k}_2{\bf k}_3{\bf k}_4}^{\Omega_1\Omega_2\Omega_3} &=& \frac{-i \Omega_1 +i \Omega_2 - i \Omega_3 + \varepsilon_{c}({\bf k}_1) - \varepsilon_{v}({\bf k}_4) -i \Omega_3 + \varepsilon_{c}({\bf k}_3) - \varepsilon_{v}({\bf k}_4)  - 2\mu }{[-i \Omega_1 + \varepsilon_{c}({\bf k}_1) - \varepsilon_{v}({\bf k}_2) - \mu][-i \Omega_2 + \varepsilon_{c}({\bf k}_3) - \varepsilon_{v}({\bf k}_2) - \mu]} \nonumber\\
    && \qquad\times\frac{1}{[-i \Omega_3 + \varepsilon_{c}({\bf k}_3) - \varepsilon_{v}({\bf k}_4) - \mu][-i \Omega_1 +i \Omega_2 + i \Omega_3 + \varepsilon_{c}({\bf k}_4) - \varepsilon_{v}({\bf k}_4) - \mu]}. \label{L_zeroT}
    \end{eqnarray}
\end{subequations}
While the fourth--order loop formula (\ref{L_zeroT}) is still quite sophisticated, within the saddle-point approximation all the denominators vanish and we get:
$$ \phi({\bf k}_1,{\bf k}_2,\Omega) = \frac{-i}{\mathcal{A}_{{\bf k}_1{\bf k}_{2}}^{\,\,\Omega}} \, \Delta({\bf k}_1,{\bf k}_2,\Omega).$$

Going back to Eq.~(\ref{BSE}), we turn to the position basis
for the relative motion ${\bf r}$ and write down the generalized Wannier equation in terms of the exciton field $\Delta({\bf r}, {\bf k}, \Omega)$:
\begin{equation}
    \left[-i \Omega + E_{\rm g} + \frac{\hbar^2 {\bf k}^2}{2 M} - \frac{\hbar^2}{2m }\nabla^2 - V(r)  - \mu \right]\Delta({\bf r}, {\bf k}, \Omega) -g_{\rm R}^2\frac{\Delta({\bf r}=0, {\bf k}, \Omega)}{E_{\rm ph}({\bf k}) -\mu -i \Omega}=0,
\end{equation}
where $m$ the the reduced mass and $M$ is the exciton mass.
It is worth noting that the same approach has been applied in Ref.~\cite{khurgin1}, albeit for the one-mode fields within the semiconductor Bloch equations approach. The field $\Delta({\bf r}, {\bf k}, \Omega)$ in such mixed representation can be 
regarded as $g_{\rm R}^2\Delta({\bf r}=0, {\bf k}, \Omega)/[E_{\rm ph}({\bf k}) -\mu -i \Omega] G_{{\bf k}, \Omega}({\bf r}, {\bf r}^\prime)$, where the Green's function is defined from the equation:
\begin{equation}\label{greens}
    \left[-i \Omega + E_{\rm g}  - \mu+ \frac{\hbar^2 {\bf k}^2}{2 M} - \frac{\hbar^2}{2m}\nabla^2 - V(r)\right]G_{{\bf k}, \Omega}({\bf r}, {\bf r}^\prime) = \delta({\bf r-r}^\prime).
\end{equation}
Here we suggest the solution in terms of the $\nu s$--state of the hydrogen-like equation, although 
the potential of interaction for $\chi_{\nu}({\bf r})$ can be considered to be of an arbitrary shape: $[- \hbar^2\nabla^2/2m - V(r)]\chi_{\nu}({\bf r}) = E_{\nu}\chi_{\nu}({\bf r})$. The Green's function then has the form:
\begin{equation}\label{greens}
    G_{{\bf k}, \Omega}({\bf r}, {\bf r}^\prime) = \sum_{\nu} A^{(\nu)}_{{\bf k}, \Omega}({\bf r'})\chi_{\nu}({\bf r}) = \sum_{\nu} \frac{\chi^{*}_{\nu}({\bf r}^\prime) \chi_{\nu}({\bf r})}{-i\Omega + \hbar^2{\bf k}^2/2M + E_{\rm g} - \mu + E_{\nu}},
\end{equation}
and the exciton field turns into
\begin{equation}
    \Delta({\bf r}, {\bf k}, \Omega) = g_{\rm R}^2\frac{\Delta({\bf r}=0, {\bf k}, \Omega)}{E_{\rm ph}({\bf k}) -\mu -i \Omega} \, G_{{\bf k}, \Omega}({\bf r}, 0).
\end{equation}
As one 
sees, the sum in the derived expression~(\ref{greens}) diverges in the point $\rm r=0$ as $\sim -\ln{r}$, as becomes clear from the form of Eq.~(\ref{greens}) when estimating the trivial limit of ${\bf k} =0$, $\Omega=0$ at $r\to0$, when one can neglect the potential compared to the delta-function in the right-hand side.
This divergence is known \cite{parish_ehph1, khurgin1} and arises because 
the attractive potential of the delta-functional form produced by the photon field leads to the reduction of the mean electron-hole separation (see e.g. \cite{khurgin}).

The exciton-photon action in momentum 
basis for the c.m. motion 
takes the form:
\begin{multline}\label{S_appB}
   \mathcal{S}= \sum_{\nu} \! \int  \!\! \frac{d {\bf k}}{(2\pi)^2} \! \int \! \frac{d \Omega}{2 \pi}\Biggl\{\bar{\Delta}({\bf r}=0, {\bf k}, \Omega)
    \left|\frac{g_{\rm R}^2}{E_{\rm ph}({\bf k})- \mu -i \Omega}\right|^2 \!\! \frac{|\chi_{\nu}({\bf r}=0)|^2}{i\Omega + E_{\rm g} - \mu + \hbar^2 {\bf k}^2/2M + E_{\nu}} \, \Delta({\bf r}=0, {\bf k}, \Omega)   \Biggr.\\[8pt]
    \Biggl. + g_{\rm R} \!\! \left[ \overline{\Psi}_{\rm ph}({\bf k}, \Omega) \frac{g_{\rm R}^2 |\chi_{\nu}({\bf r}=0)|^2}{E_{\rm ph}({\bf k}) \!-\! \mu \!-\! i \Omega} \, \frac{\Delta({\bf r} = 0, {\bf k}, \Omega)}{-i\Omega \!+\! E_{\rm g} \!-\! \mu \!+\! \hbar^2 {\bf k}^2/2M \!+\! E_{\nu}} \!+\! \frac{\bar{\Delta}({\bf r} =0, {\bf k}, \Omega)}{{i\Omega \!+\! E_{\rm g} \!-\! \mu \!+\! \hbar^2{\bf k}^2/2M \!+\! E_{\nu}}} \, \frac{g_{\rm R}^2  |\chi_{\nu}({\bf r} = 0)|^2}{E_{\rm ph}({\bf k}) \!-\! \mu \!+\! i \Omega}\Psi_{\rm ph}({\bf k}, \Omega) \!\right]\!\Biggr\} \\[8pt]
    \hspace{250pt} + \int \!\! \frac{d {\bf k}}{(2\pi)^2} \! \int \! \frac{d \Omega}{2 \pi} \,\, \overline{\Psi}_{\rm ph}({\bf k},\Omega)\bigl[- i \Omega + E_{\rm ph}({\bf k}) -\mu\bigr] \Psi_{\rm ph}({\bf k},\Omega)\\[8pt]
    + \frac{1}{2} \! \int \!\! \frac{d\bf p}{(2\pi)^2} \! \int \! \frac{d{\bf l}_1 d{\bf l}_2 d{\bf l}_3}{(2\pi)^6} \! \int \! \frac{d\Omega_1d\Omega_2d\Omega_3}{(2\pi)^3} \left[ -i \Omega_1 + i \Omega_2 - i \Omega_3 + \varepsilon_{c}\Bigl({\bf p} \!+\! \frac{{\bf l}_1}{2}\Bigr) - \varepsilon_{v}\Bigl({\bf p} \!+\! {\bf l}_2 \!-\! {\bf l}_3 \!-\! \frac{{\bf l}_1}{2}\Bigr) - i \Omega_3 + \varepsilon_{c}\Bigl({\bf p} \!+\! {\bf l}_{2} \!-\! \frac{{\bf l}_{1}}{2}\Bigr) \right.\\[8pt]
    \left.- \varepsilon_{v}\Bigl({\bf p} \!+\! {\bf l}_{2} \!-\! {\bf l}_3 \!-\! \frac{{\bf l}_{1}}{2}\Bigr)\right]  \Delta({\bf p}, {\bf l}_1, \Omega_1)\bar{\Delta}\Bigl({\bf  p} \!+\! \frac{{\bf l}_2 \!-\! {\bf l}_1}{2}, {\bf l}_2, \Omega_2\Bigr)
    \Delta\Bigl({\bf p} \!+\! {\bf l}_2 \!-\!\frac{{\bf l}_3\!+\!{\bf l}_1}{2}, {\bf l}_3, \Omega_3\Bigr) \bar{\Delta}\Bigl( {\bf p} \!+\! \frac{{\bf l}_2 \!-\! {\bf l}_3}{2}, {\bf l}_1 \!-\! {\bf l}_2 \!+\! {\bf l}_3, \Omega_1 \!-\! \Omega_2 \!+\! \Omega_3\Bigr).
\end{multline}
As one can see from three first terms in Eq.~(\ref{S_appB}), it is natural to introduce 
a field corresponding to the exciton as a whole (i.e. depending only on the total momentum):
$$C^{(\nu)}({\bf k},\Omega) =  \frac{g_{\rm R}^2 \chi^{*}_{\nu}({\bf r} =0)}{E_{\rm ph}({\bf k})- \mu -i \Omega} \frac{\Delta({\bf r}=0, {\bf k}, \Omega)}{[-i\Omega + E_{\rm g}  - \mu + \hbar^2{\bf k}^2/2M + E_{\nu}]},$$
hence the exciton field $\Delta$ can be rewritten as
$$ \Delta({\bf p}, {\bf k}, \Omega) = \sum_{\nu} C^{(\nu)}\!({\bf k}, \Omega) \chi_{\nu}({\bf p}).$$
Provided all these simplifications and with the exciton interaction term taken in the saddle point according to Eq.~(\ref{saddle-point-eqA}), the action (\ref{S_appB}) takes the form
\begin{multline}
     \mathcal{S}= \sum_{\nu} \! \int \!\! \frac{d {\bf k}}{(2\pi)^2} \int \! \frac{d \Omega}{2 \pi} \left\{ \overline{C}^{(\nu)}\!({\bf k}, \Omega)\!\left(-i\Omega \!+\! E_{\rm g} \!-\! \mu \!+\! \frac{\hbar^2{\bf k}^2}{2 M} \!+\! E_{\nu}\right)\!C^{(\nu)}\!({\bf k}, \Omega) + g_{\rm R} \!\left[\overline{C}^{(\nu)}\!({\bf k}, \Omega)\chi^{*}_{\nu}({\bf r}=0)  \Psi_{\rm ph}({\bf k}, \Omega) \right.\right.
     \\[6pt]
     \biggl.\left. + \overline{\Psi}_{\rm ph}({\bf k}, \Omega) C^{(\nu)}\!({\bf k}, \Omega) \chi_{\nu}({\bf r}=0) \right]\biggr\} + \int \!\! \frac{d {\bf k}}{(2\pi)^2} \int \! \frac{d \Omega}{2 \pi} \,\, \overline{\Psi}_{\rm ph}({\bf k},\Omega)\bigl[- i \Omega + E_{\rm ph}({\bf k}) -\mu\bigr]\Psi_{\rm ph}({\bf k},\Omega)
     \\[6pt]
     +\! \frac{1}{2} \!\! \sum_{\nu_1\dots\nu_4} \!\! \int\!\! \frac{d{\bf p}}{(2\pi)^2} \!\! \int \!\frac{d{\bf l}_1 d{\bf l}_2 d{\bf l}_3}{(2\pi)^6} \!\!\! \int \! \frac{d\Omega_1d\Omega_2d\Omega_3}{(2\pi)^3} C^{(\nu_1)}\!({\bf l}_1, \Omega_1) \chi_{\nu_1}\!({\bf p})\overline{C}^{(\nu_2)}\!({\bf l}_2, \Omega_2) \chi^{*}_{\nu_2}\!\Bigl({\bf p} + \frac{{\bf l}_2 \!-\! {\bf l}_1}{2}\!\Bigr)C^{(\nu_3)}\!({\bf l}_3, \Omega_3) \chi_{\nu_3}\!\Bigl({\bf p} + {\bf l}_2 - \frac{{\bf l}_3 \!+\! {\bf l}_1}{2}\!\Bigr) \\[6pt]
     \times
    2 \! \int \!\! \frac{d{\bf q}}{(2\pi)^2} \, V({\bf q}) \, \overline{C}^{(\nu_4)}\!({\bf l}_1 \!-\! {\bf l}_2 \!+\! {\bf l}_3, \Omega_1 \!-\! \Omega_2 \!+\! \Omega_3 )\chi^{*}_{\nu_4}\!\Bigl({\bf p} \!+\! \frac{{\bf l}_2 \!-\! {\bf l}_3}{2} \!-\! {\bf q}\Bigr)
    \\[6pt]
    -\frac{1}{2} \sum_{\nu_1\dots\nu_3} \! \int \!\! \frac{d{\bf p}}{(2\pi)^2} \! \int \!\frac{d{\bf l}_1 d{\bf l}_2 d{\bf l}_3}{(2\pi)^6} \! \int \! \frac{d\Omega_1d\Omega_2d\Omega_3}{(2\pi)^3} \, \left[ C^{(\nu_1)}\!({\bf l}_1, \Omega_1) \chi_{\nu_1}\!({\bf p})\overline{C}^{(\nu_2)}\!({\bf l}_2, \Omega_2) \chi^{*}_{\nu_2}\!\Bigl({\bf p}\!+\! \frac{{\bf l}_2 \!-\! {\bf l}_1}{2}\Bigr) \right. \\ \biggl. \times  C^{(\nu_3)}\!({\bf l}_3, \Omega_3) \chi_{\nu_3}\!\Bigl({\bf p} \!+\! {\bf l}_2 \!-\! \frac{{\bf l}_3 \!+\!{\bf l}_1}{2}\Bigr) g_{\rm R}\overline{\Psi}_{\rm ph}({\bf l}_1- {\bf l}_2+{\bf l}_3, \Omega_1 -\Omega_2 +\Omega_3 ) + \text{c.c.}\biggr], \label{action_B}
\end{multline}
where the last term describes the so-called ``saturation'' or, 
to be precise, the exciton-assisted exciton-photon coupling. 
We note here that to the final expression for action one should add the contribution to the exciton-exciton interaction arising from the electron screening derived in Appendix~\ref{appA}:
\begin{multline}\label{add_scr}
   \mathcal{S}^\prime_{\rm scr} = - \frac{1}{2} \sum_{\nu_1\dots\nu_4} \! \int \!\! \frac{d{\bf p}}{(2\pi)^2} \! \int \!\frac{d{\bf l}_1 d{\bf l}_2 d{\bf l}_3}{(2\pi)^6} \! \int \! \frac{d\Omega_1d\Omega_2d\Omega_3}{(2\pi)^3} \, C^{(\nu_1)}\!({\bf l}_1, \Omega_1)\chi_{\nu_1}\!({\bf p})\overline{C}^{(\nu_2)}\!({\bf l}_2, \Omega_2) \chi^{*}_{\nu_2}\Bigl({\bf p} \!+\! \frac{{\bf l}_2 \!-\! {\bf l}_1}{2}\Bigr) \\[5pt]
   \times 2 \! \int \!\! \frac{d{\bf q}}{(2\pi)^2} \, V({\bf q})C^{(\nu_3)}\!({\bf l}_3, \Omega_3) \chi_{\nu_3}\!\Bigl({\bf p \!-\! q}  \!+\! {\bf l}_2 \!-\! \frac{{\bf l}_3 \!+\! {\bf l}_1}{2}\Bigr) \overline{C}^{(\nu_4)}\!({\bf l}_1 \!-\! {\bf l}_2 \!+\! {\bf l}_3, \Omega_1 \!-\! \Omega_2 \!+\! \Omega_3)\chi^{*}_{\nu_4}\!\Bigl({\bf p \!-\! q} \!+\! \frac{{\bf l}_2\!-\! {\bf l}_3}{2}\Bigr).
\end{multline}
The action obtained as a sum of Eqs.~(\ref{action_B}) and (\ref{add_scr}) has the form quite similar to the commonly-used action, but containing the sums over the exciton $s$--states.

\section{The ``rigid'' exciton limit}\label{appC}

In the following we restrict ourselves with the standard case in which only the $1s$--exciton state is assumed to appear in the system, revealing how the expressions for interactions can be transformed into the widely known values of interaction constants.  
To this end, it is convenient to restrict ourselves in Eq.~(\ref{BSE}) to $\chi^{(1)}_{{\bf k},\Omega}({\bf p})\equiv \chi_{\bf k}({\bf p})$ and the strong-coupling regime $\hbar\Omega_{\rm R} \ll E_{\rm b}$, and thus neglect the term $\sim(\hbar \Omega_{\rm R})^2/E_{\rm ph}$. In this case, without the van-der-Waals term (see Appendix~\ref{appA}) Eq.~(\ref{BSE}) turns into:
\begin{multline}\label{D0}
    \left(E_{\rm g} +\frac{\hbar^2 {\bf k}^2}{2 M} +\frac{\hbar^2 {\bf p}^2 }{2m} - \mu - i \Omega\right)\Delta({\bf p},{\bf k}, \Omega) -\! \left[n_{v}\!\left(\!{\bf p} - \!\frac{m}{m_v} {\bf k}\!\right)\! - n_{c}\!\left(\!{\bf p}\! + \!\frac{m}{m_c} {\bf k}\!\right)\!\right]\int \!\!\frac{d{\bf q}}{(2\pi)^2}V({\bf p} -{\bf q}) \Delta( {\bf q},{\bf k},\Omega) =0
\end{multline}
and can be solved using the ansatz $\Delta( {\bf p},{\bf k},\Omega) = \chi_{\bf k}({\bf p})C({\bf k}, \Omega)$ where the $1s$ exciton wavefunction $\chi_{\bf k}({\bf p})$ is modified by temperature.

At $T=0$, the saddle-point equation for configuration of the field $\Delta$ turns into the standard Wannier equation:
\begin{equation}
   \left(E_{\rm g} +\frac{\hbar^2 {\bf k}^2}{2 M} +\frac{\hbar^2 {\bf p}^2 }{2m} - \mu - i \Omega\right)\Delta({\bf p},{\bf k}, \Omega) -\int \frac{d{\bf q}}{(2\pi)^2}V({\bf p} -{\bf q}) \Delta( {\bf q},{\bf k},\Omega) =0.
\end{equation}
The solution is  $\Delta({\bf p}, {\bf k}, \Omega) = \chi({\bf p})C({\bf k}, \Omega)$ where $\chi({\bf p})$ is the $1s$--exciton  wavefunction and the excited $s$--states are neglected:
$$\left(\frac{{\bf p}^2}{2m} - E_{\rm 1s}\right)\chi({\bf p}) = \int \frac{d{\bf q}}{(2\pi)^2}V({\bf p} -{\bf q})\chi({\bf q}).$$
In this case, the saddle-point equations (\ref{saddle-point-eqA}) and (\ref{saddle-point-eqB}) reduce to the standard Hopfield equations, and  the exciton-photon action takes the well-known form \cite{lozovik_PI}:
\begin{multline}\label{action_rigid}
     \mathcal{S}= \!\int \!\! \frac{d {\bf k}}{(2\pi)^2} \! \int \!\! \frac{d \Omega}{2 \pi} \left\{\overline{C}({\bf k}, \Omega) \! \left(\!-i\Omega + E_{\rm g}  - \mu + \frac{\hbar^2 {\bf k}^2}{2 M} + E_{\rm 1s} \!\right)C({\bf k}, \Omega)+  \overline{\Psi}_{\rm ph}({\bf k},\Omega)(- i \Omega + E_{\rm ph}({\bf k}) -\mu)\Psi_{\rm ph}({\bf k},\Omega) \right.\\[5pt]
     \hspace{220pt}
     \biggl.+\frac{\hbar\Omega_{\rm R}}{2} \Bigl[\overline{C}({\bf k}, \Omega)\Psi_{\rm ph}({\bf k}, \Omega)  + \overline{\Psi}_{\rm ph}({\bf k}, \Omega) C({\bf k}, \Omega) \Bigr] \biggr\}
     \\[5pt]
     + \frac{1}{2} \!\int\!\frac{d{\bf l}_1 d{\bf l}_2 d{\bf l}_3}{(2\pi)^6} \! \int \! \frac{d\Omega_1d\Omega_2d\Omega_3}{(2\pi)^3} \Bigl\{V_{\rm ex}^{tot}({\bf l}_1,{\bf l}_2, {\bf l}_3)  C({\bf l}_1, \Omega_1) \overline{C}({\bf l}_2, \Omega_2) C({\bf l}_3, \Omega_3)\overline{C}({\bf l}_1 \!-\! {\bf l}_2 \!+\! {\bf l}_3, \Omega_1 \!-\! \Omega_2 \!+\! \Omega_3) \Bigr.\\[5pt]
     \!-\!  
     \left. \Bigl[V_{\rm sat}({\bf l}_1, {\bf l}_2, {\bf l}_3)C({\bf l}_1, \Omega_1) \overline{C}({\bf l}_2, \Omega_2) C({\bf l}_3, \Omega_3)\overline{\Psi}_{\rm ph}({\bf l}_1 \!-\! {\bf l}_2 \!+\! {\bf l}_3, \Omega_1 \!-\! \Omega_2 \!+\! \Omega_3)  + \text{c.c.}\Bigr]\!\right\}
\end{multline}
with $\hbar \Omega_{\rm R}/2 = g_{\rm R} \sum_{{\bf q}}\chi({\bf q}) = g_{\rm R} \chi({\bf r}=0)$ being the experimentally relevant Rabi-splitting (whereas $g_{\rm R}$ is the bare electron-hole-photon coupling rate), and the notations
\begin{align}\label{Vtot}
     V^{tot}_{\rm ex}({\bf l}_1,{\bf l}_2,{\bf l}_3) & \!\equiv 2 \!\!\int \!\! \frac{d{\bf p}}{(2\pi)^2} \!\! \int \!\! \frac{d{\bf q}}{(2\pi)^2} V({\bf q}) \chi({\bf p}) \bar{\chi}\Bigl({\bf p} \!+\! \frac{{\bf l}_2 \!-\! {\bf l}_1}{2}\Bigr)\bar{\chi}\Bigl({\bf p} \!+\! \frac{{\bf l}_2 \!-\! {\bf l}_3}{2} \!-\!{\bf q}\Bigr)\!\biggl[\chi\Bigl({\bf p} \!+\! {\bf l}_2 \!-\! \frac{{\bf l}_3 \!+\! {\bf l}_1}{2}\Bigr) \!-\! \chi\Bigl({\bf p} \!+\! {\bf l}_2 \!-\! \frac{{\bf l}_3 \!+\! {\bf l}_1}{2} \!-\!{\bf q}\Bigr)\!\biggr]
     \\
    \label{Vsat}
    V_{\rm sat}({\bf l}_1,{\bf l}_2,{\bf l}_3) & \equiv g_{\rm R} \!\int \!\! \frac{d{\bf p}}{(2\pi)^2} \, \chi({\bf p})\bar{\chi}\Bigl({\bf p}\!+\!\frac{{\bf l}_2 \!-\! {\bf l}_1}{2}\Bigr) \chi\Bigl({\bf p} \!+\! {\bf l}_2 \!-\! \frac{{\bf l}_3 \!+\! {\bf l}_1}{2}\Bigr).
\end{align}
One sees that depending on the material and the interaction potential between the charges within the exciton, the shape of $\chi({\bf p})$ may vary thus alternating both the exciton-exciton (interaction) and the saturation nonlinearities in the polariton system. 
%
We note as well that the exciton-exciton interaction term including Eq.~(\ref{Vtot}) is usually calculated 
in the operator formalism~\cite{yamamoto, glazov, parish_ehph1,combescot, trion-polaritons}. Here we stress that within our approach we derive excitonic rather than polaritonic interactions (if one instead solves the Bethe-Salpeter equation modified by light (\ref{BSE}), the obtained interactions of the renormalized field would be ``polaritonic'').

Finally, assuming that 
the c.m. momenta ${\bf l}_i$ ($i=1,2,3$) are negligibly small compared to the momentum of internal exciton motion, the  exciton-exciton interaction constant obtained from Eq.~(\ref{Vtot}) coincides with the well-known result:
\begin{equation}\label{gex}
     g_{\rm ex}= 2 \int \! \frac{d{\bf p}}{(2\pi)^2}  \int \! \frac{d{\bf q}}{(2\pi)^2} V({\bf q})\chi({\bf p})\chi^{*}({\bf p})\chi^{*}({\bf p} -{\bf q})\bigl[\chi({\bf p} )-\chi({\bf p -q} )\bigr].
\end{equation}
The 
saturation interaction constant is obtained similarly from Eq.~(\ref{Vsat}): 
\begin{equation}\label{gsat}
     g_{\rm sat} = \frac{\hbar \Omega_{\rm R}}{2\chi({\bf r}=0)} \int \! \frac{d{\bf p}}{(2\pi)^2}\, \chi{(\bf p)}\chi^{*}{(\bf p)}\chi{(\bf p)},
\end{equation}
which coincides with the expressions derived using alternative approaches in Refs.~\cite{yamamoto, schwendimann, glazov, binder_scat}.

Starting from the action (\ref{action_rigid}) 
and considering 
the limit of Eqs.~(\ref{gex}),~(\ref{gsat}),
\begin{multline}
     \mathcal{S} = \! \int \!\! \frac{d {\bf k}}{(2\pi)^2} \! \int \! \frac{d \Omega}{2 \pi} \! \left\{\overline{C}({\bf k}, \Omega)\!\left(-i\Omega + E_{\rm g}  - \mu + \frac{\hbar^2 {\bf k}^2}{2 M} + E_{\rm 1s} \!\right)C({\bf k}, \Omega) +  \overline{\Psi}_{\rm ph}({\bf k},\Omega)\bigl[- i \Omega + E_{\rm ph}({\bf k}) -\mu\bigr]\Psi_{\rm ph}({\bf k},\Omega) +\right.\\[5pt]
     \left. + \frac{\hbar\Omega_{\rm R}}{2} \Bigl[\overline{C}({\bf k}, \Omega)\Psi_{\rm ph}({\bf k}, \Omega)  + \overline{\Psi}_{\rm ph}({\bf k}, \Omega) C({\bf k}, \Omega) \Bigr] \right\}
     + \! \int \!\! \frac{d{\bf l}_1d{\bf l}_2d{\bf l}_3}{(2\pi)^6} \!\! \int \! \frac{d\Omega_1d\Omega_2d\Omega_3}{(2\pi)^3} \left\{\frac{g_{\rm ex}}{2}C({\bf l}_1, \Omega_1) \overline{C}({\bf l}_2, \Omega_2)  C({\bf l}_3, \Omega_3) \right. \\[5pt]
     \times\overline{C}({\bf l}_1- {\bf l}_2+{\bf l}_3, \Omega_1 -\Omega_2 +\Omega_3) \left. -  \frac{g_{\rm sat}}{2} 
     \left[ C({\bf l}_1, \Omega_1) \overline{C}({\bf l}_2, \Omega_2)  C({\bf l}_3, \Omega_3)\overline{\Psi}_{\rm ph}({\bf l}_1- {\bf l}_2+{\bf l}_3, \Omega_1 -\Omega_2 +\Omega_3 ) + {\rm c.c.}\right]\right\},
\end{multline}
we perform the saddle-point approximation assuming the system to be in equilibrium $i\Omega = 0$:
\begin{subequations}
    \begin{eqnarray}
       \left. \frac{\delta {\mathcal{S}}_{\rm eff}}{\delta \overline{C}} \right|_{
       \substack{C=\Psi_{\rm ex} \\ \Psi_{\rm ph}\quad}} &=& (E_{\rm g} + E_{1s}-\mu) \Psi_{\rm ex} + \frac{\hbar\Omega_{\rm R}}{2}\Psi_{\rm ph} +g_{\rm ex} |\Psi_{\rm ex}|^2 \Psi_{\rm ex} -g_{\rm sat} |\Psi_{\rm ex}|^2 \Psi_{\rm ph} - \frac{g_{\rm sat}}{2}(\Psi_{\rm ex})^2 \overline{\Psi}_{\rm ph}= 0, \\
       \left. \frac{\delta {\mathcal{S}}_{\rm eff}}{\overline{\Psi}_{\rm ph}^0} \right|_{
       \substack{C=\Psi_{\rm ex} \\ \Psi_{\rm ph}\quad}} &=& (E_{\rm ph}^0-\mu) \Psi_{\rm ph} +\frac{\hbar\Omega_{\rm R}}{2} \Psi_{\rm ex} - \frac{g_{\rm sat}}{2} |\Psi_{\rm ex}|^2 \Psi_{\rm ex}=0
    \end{eqnarray}
\end{subequations}
where $\Psi_{\rm ex(ph)}$ are the equilibrium exciton (photon) fields. 
The relation between the two 
can be defined as:
\begin{equation}
    \Psi_{\rm ph} = \frac{\Psi_{\rm ex}}{\mu  - E_{\rm ph}^0} \left(\frac{\hbar \Omega_{\rm R}}{2} - \frac{g_{\rm sat}|\Psi_{\rm ex}|^2}{2}\right),
\end{equation}
and the phase difference between 
them being $0$ or $\pi$ depending on $\mu$. Solving these algebraic equations, one can calculate the bottoms of dispersion laws of the two new normal modes, i.e. lower (LP) and upper (UP) polaritons, which depend on the uniform exciton density $n_{\rm ex}^0$:
\begin{equation}
    \mu_{\rm LP(UP)}(n_{\rm ex}^0) = \frac{E_{\rm g} + E_{1s} + E_{\rm ph}^{0}+ g_{\rm ex} n_{\rm ex}^0}{2} \mp \frac{1}{2}\sqrt{( g_{\rm ex} n_{\rm ex}^0 - \Delta)^2 + (\hbar \Omega_{\rm R} - g_{\rm sat} n_{\rm ex}^0)(\hbar \Omega_{\rm R} - 3 g_{\rm sat} n_{\rm ex}^0)},
\end{equation}
where $E_{\rm ph}^0 - (E_{\rm g} + E_{1s})= \Delta$ is usually regarded as the (constant) energy detuning between photons and excitons at ${\bf k}=0$.

Rewriting the formulae in terms of the renormalized (density-dependent) detuning $\Delta \to \Delta(n_{\rm ex}^0) = \Delta - g_{\rm ex} n_{\rm ex}^0$ and Rabi-splitting $\hbar \Omega_R \to \hbar \Omega_R(n_{\rm ex}^0) = \sqrt{(\hbar \Omega_{\rm R} - g_{\rm sat} n_{\rm ex}^0)(\hbar \Omega_{\rm R} - 3 g_{\rm sat} n_{\rm ex}^0)}$, we get:
\begin{equation}
     \mu_{\rm LP(UP)}(n_{\rm ex}^0) = E_{\rm g} + E_{1s}+ g_{\rm ex} n_{\rm ex}^0 + \frac{\Delta (n_{\rm ex}^0)}{2} \mp\frac{1}{2}\sqrt{\Delta^2(n_{\rm ex}^0) + [\hbar\Omega_R(n_{\rm ex}^0)]^2}.
\end{equation}
One sees that the separation of the bottoms of dispersions (at ${\bf k}=0$) shifts with the exciton density as
\begin{equation}
    \mu_{\rm UP}(n_{\rm ex}^0)- \mu_{\rm LP}(n_{\rm ex}^0) = \sqrt{\Delta^2(n_{\rm ex}^0) + [\hbar\Omega_R(n_{\rm ex}^0)]^2}.
\end{equation}
At the same time, from the experimental point of view a more relevant quantity characterizing the exciton-photon conversion rate is the distance between the polariton branches at the anticrossing point (which is shifting itself with the change of the exciton density, since the detuning is also changing). It can be defined as follows: \begin{equation}
  \min[E_{\rm UP}({\bf k})- E_{\rm LP}({\bf k})] = \sqrt{\Delta_{\bf k}^2(n_{\rm ex}^0) + [\hbar\Omega_R(n_{\rm ex}^0)]^2} \;\,\text{with} \; \Delta_{\bf k}(n_{\rm ex}^0) = \Delta - g_{\rm ex} n_{\rm ex}^0 + \frac{\hbar^2 {\bf k}^2}{2 m_{\rm ph}} - \frac{\hbar^2 {\bf k}^2}{2 m_{\rm ex}}.
\end{equation}

\section{Consideration of spins}\label{appD}
For purely excitonic systems, the treatment of spins within functional integration approach was briefly mentioned in Ref.~\cite{tokatly} without any detailed discussion. The aim of this Appendix is to study the influence of spins on the effective exciton-photon action in polariton systems.
We restrict ourselves to the simplest case, considering $T=0$ and rigid  $1s$--excitons. Moreover, we will focus on the corrections to the expressions derived above arising due to {\it dark exciton states}, so the density channels of pairing addressed in Appendix~\ref{appA} at this stage are neglected for convenience. Below we shortly discuss the screening of exciton interaction for excitons with different spins.
Even such a simple case upon examination turns out to be intricate.

Our starting point is the action without spin-orbit coupling:
\begin{multline}\label{action_e-h-ph_initial_spin}
    \mathcal{S}[\Psi_{\!c}, \Psi_{\!v}, \Psi_{\rm ph}] = \!\int \!\!d{\bf r} \! \int\limits_{0}^{\beta} \!\! d\tau \! \left\{ \begin{pmatrix}
             \overline{\Psi}_{\!c \uparrow}(x) \\ \overline{\Psi}_{\!c\downarrow}(x) \\\overline{\Psi}_{\!v\uparrow}(x) \\ \overline{\Psi}_{\!v \downarrow}(x)
        \end{pmatrix}^{\!\!\! T} \!\!\!
        \begin{pmatrix}
            \partial_{\tau} + \varepsilon_{c \uparrow}({\bf\hat{k}}) & 0 & 0  & 0 \\ 0 &  \partial_{\tau} + \varepsilon_{c\downarrow}({\bf\hat{k}}) & 0 & 0\\ 0 & 0 &  \partial_{\tau} + \varepsilon_{v\uparrow}({\bf\hat{k}}) & 0 \\ 0 & 0  & 0& \partial_{\tau} + \varepsilon_{v \downarrow}({\bf\hat{k}})
        \end{pmatrix} \!\!
        \begin{pmatrix}
             {\Psi}_{\!c \uparrow}(x) \\  {\Psi}_{\!c \downarrow}(x) \\ {\Psi}_{\!v \uparrow}(x) \\ {\Psi}_{\!v \downarrow}(x)
        \end{pmatrix} \right.
        \\
        \Biggl. + g_{\rm R} \bigl[\overline{\Psi}_{\rm ph \uparrow}(x)\Psi_{\!c \downarrow}(x)\overline{\Psi}_{\!v \uparrow}(x) +\overline{\Psi}_{\!c \downarrow}(x)\Psi_{\!v \uparrow}(x)\Psi_{\rm ph\uparrow}(x) + \overline{\Psi}_{ph \downarrow}(x)\Psi_{\!c \uparrow}(x)\overline{\Psi}_{\!v \downarrow}(x) +\overline{\Psi}_{\!c \uparrow}(x)\Psi_{\!v \downarrow}(x)\Psi_{\rm ph\downarrow}(x) \Bigr]\Biggr\}
        \\[-2pt]
        +  \sum_{\sigma}\!\int \!\! d{\bf r}\! \int\limits_{0}^{\beta} \!\! d\tau \,\overline{\Psi}_{\rm ph, \sigma}(x)\Bigl[\partial_{\tau} + E_{\rm ph \sigma}({\bf\hat{k}})\Bigr] \Psi_{\rm ph, \sigma}(x)
        +\frac{1}{2}\sum_{i, j} \sum_{\sigma, \sigma^\prime} \! \int\!\! d{\bf r}d{\bf r^\prime}\!\int\limits_{0}^{\beta} \!\! d\tau d\tau^\prime V(x-x^\prime) \overline{\Psi}_{\!i \sigma}(x)\Psi_{\!i \sigma}(x) \overline{\Psi}_{\!j \sigma^\prime}(x^\prime)\Psi_{\!j \sigma^\prime}(x^\prime),
\end{multline}
where $\sigma = \uparrow, \downarrow$ is a spin index, $x=({\bf r},t)$ and $i,j=c,v$ as before, $\varepsilon_{c(v) \uparrow (\downarrow)}({\bf k}) = \pm E_{\rm g}/2 \pm \hbar^2 {\bf k}^2/2m_{c(v)\uparrow (\downarrow)} - \mu_{c(v)\uparrow (\downarrow)}$, $E_{\rm ph \uparrow(\downarrow)}({\bf k}) = E_{\rm ph}^0 + \hbar^2 {\bf k}^2/2 m_{\rm ph} -\mu_{\rm ph \uparrow (\downarrow)}$. Here, in general case, we assume the chemical potentials and effective masses of electrons with different spins in conduction and valence bands being different.
We introduce four exciton fields corresponding to the bright and dark excitons. For TMD materials, where the electrons in both valence and conduction bands have spin projections $\pm1/2$, one has:
    \begin{equation}
        \Delta_{+1}(x, y) = \overline{\Psi}_{\!v\uparrow} (y)\Psi_{\!c \uparrow}(x),\,  \Delta_{-1}(x, y) = \overline{\Psi}_{\!v\downarrow} (y)\Psi_{\!c \downarrow}(x),\,  \Delta_{0}(x, y) = \overline{\Psi}_{\!v\uparrow} (y)\Psi_{\!c \downarrow}(x), \ \Delta_{\bar{0}}(x, y) = \overline{\Psi}_{\!v\downarrow} (y)\Psi_{\!c \uparrow}(x),
    \end{equation}
while for GaAs, where we consider only heavy holes, i.e. electrons in the valence band with spin projections $\pm 3/2$ which 
correspond to $\Psi_{\!v \uparrow}$ and $\Psi_{\!v \downarrow}$, respectively,
    \begin{equation}
        \Delta_{+1}(x, y) = \overline{\Psi}_{\!v\uparrow} (y)\Psi_{\!c \downarrow}(x),\,  \Delta_{-1}(x, y) = \overline{\Psi}_{\!v\downarrow} (y)\Psi_{\!c \uparrow}(x),\,  \Delta_{+2}(x, y) = \overline{\Psi}_{\!v\uparrow} (y)\Psi_{\!c \uparrow}(x), \ \Delta_{-2}(x, y) = \overline{\Psi}_{\!v\downarrow} (y)\Psi_{\!c \downarrow}(x).
    \end{equation}
In the following, we will not make a distinction between the materials and will generically denote the two dark fields as $\Delta_{\rm d,\bar{d}}$. Along with the four exciton fields and their conjugates, in the same way as in the spinless case (using 8 $\delta$--functions, see the main text), we have 4 auxiliary fields $\phi_{\pm1}$, $\phi_{\rm d,\bar{d}}$ and their conjugates. 
The action (\ref{action_e-h-ph_initial_spin}) takes the form:
\begin{multline}\label{action_e-h-ph_spin}
    {\mathcal{S}}[\Psi_{\!c,v}, \Psi_{\rm ph}, \phi_{\pm1,\rm d,\bar{d}}, \Delta_{\pm1,\rm d,\bar{d}}] = \! \sum_{{\bf k},\omega} \! \begin{pmatrix}
             \overline{\Psi}_{\!c \uparrow}(k) \\ \overline{\Psi}_{\!c \downarrow}(k) \\\overline{\Psi}_{\!v \uparrow}(k) \\ \overline{\Psi}_{\!v  \downarrow}(k)
        \end{pmatrix}^{\!\!\!T} \!\!\!
        \begin{pmatrix}
            -i \omega + \varepsilon_{c \uparrow}({\bf{k}}) & 0 & 0  & 0 \\ 0 &  -i \omega+ \varepsilon_{c\downarrow}({\bf{k}}) & 0 & 0\\ 0 & 0 &   -i \omega + \varepsilon_{v\uparrow}({\bf k}) & 0 \\ 0 & 0  & 0&  -i \omega + \varepsilon_{v \downarrow}({\bf k})
        \end{pmatrix} \!\!
        \begin{pmatrix}
             {\Psi}_{\!c  \uparrow}(k) \\  {\Psi}_{\!c  \downarrow}(k) \\ {\Psi}_{\!v  \uparrow}(k) \\ {\Psi}_{\!v  \downarrow}(k)
        \end{pmatrix}
        \\[5pt]
    + i\sqrt{T} \!\! \sum_{\substack{{\bf k}_1,{\bf k}_2 \\ \omega_1, \omega_2}} \!
        \begin{pmatrix}
             \overline{\Psi}_{\!c \uparrow}(k_1) \\ \overline{\Psi}_{\!c \downarrow}(k_1) \\\overline{\Psi}_{\!v \uparrow}(k_1) \\ \overline{\Psi}_{\!v \downarrow}(k_1)
        \end{pmatrix}^{\!\!\!T} \!\!
       \delta\mathcal{G}^{-1}(k_1, k_2) \!\!
        \begin{pmatrix}
             {\Psi}_{\!c  \uparrow}(k_2) \\  {\Psi}_{\!c  \downarrow}(k_2) \\ {\Psi}_{\!v  \uparrow}(k_2) \\ {\Psi}_{\!v  \downarrow}(k_2)
        \end{pmatrix}
        -i \!\! \sum_{\substack{\sigma = \pm1, \\ \rm d,\bar{d}}}\sum_{\substack{{\bf k}_1,{\bf k}_2 \\ \Omega}}\! \Bigl[ \bar{\phi}_{\sigma}({\bf k}_1,{\bf k}_2, \Omega)\Delta_{\sigma}({\bf k}_1,{\bf k}_2, \Omega) + \bar{\Delta}_{\sigma}({\bf k}_1,{\bf k}_2, \Omega) \phi_{\sigma}({\bf k}_1,{\bf k}_2, \Omega) \Bigr]
        \\[5pt]
        - \!\! \sum_{\substack{\sigma = \pm1, \\ \rm d,\bar{d}}} \sum_{\substack{{\bf k}_1\dots{\bf k}_4 \\ \Omega}} \!\! V({\bf k}_1 - {\bf k}_2) \bar{\Delta}_{\sigma}({\bf k}_1,{\bf k}_4,\Omega) \Delta_{\sigma}({\bf k}_2,{\bf k}_3,\Omega) \delta({\bf k}_1+{\bf k}_3, {\bf k}_2+{\bf k}_4) + \!\! \sum_{\sigma = \pm1}\sum_{{\bf k}, \Omega} \overline{\Psi}_{\rm ph,\sigma}(k)\bigl[- i \Omega + E_{\rm ph}({\bf k}) -\mu\bigr]\Psi_{\rm ph,\sigma}(k)
        \\
        + \frac{g_{\rm R}}{\sqrt{S}} \sum_{\sigma = \pm1}\sum_{\substack{{\bf k}_1,{\bf k}_2 \\ \Omega}} \Bigl[\overline{\Psi}_{\rm ph,\sigma}({\bf k}_1-{\bf k}_2,\Omega) \Delta_{\sigma}({\bf k}_1,{\bf k}_2,\Omega) + \bar{\Delta}_{\sigma}({\bf k}_1,{\bf k}_2,\Omega)\Psi_{\rm ph, \sigma}({\bf k}_1-{\bf k}_2,\Omega)\Bigr],
\end{multline}
where
\begin{equation}\label{deltaG}
    \delta\mathcal{G}^{-1}(k_1, k_2)= \begin{pmatrix}
            0 & 0  & \phi_{\rm d} ({\bf k}_1,{\bf k}_2, \omega_1-\omega_2)& \phi_{-1} ({\bf k}_1,{\bf k}_2,  \omega_1-\omega_2)\\   0 & 0  & \phi_{+1} ({\bf k}_1,{\bf k}_2,  \omega_1-\omega_2)& \phi_{\bar{\rm d}} ({\bf k}_1,{\bf k}_2,  \omega_1-\omega_2)  \\ \bar{\phi}_{\rm d} ({\bf k}_2,{\bf k}_1,  \omega_2-\omega_1)& \bar{\phi}_{+1} ({\bf k}_2,{\bf k}_1, \omega_2-\omega_1) & 0 & 0 \\ \bar{\phi}_{-1} ({\bf k}_2,{\bf k}_1, \omega_2-\omega_1)& \bar{\phi}_{\bar{\rm d}} ({\bf k}_2,{\bf k}_1, \omega_2-\omega_1) & 0 & 0
        \end{pmatrix}.
\end{equation}
In the above expressions, we used the spin composition for GaAs, but the result is similarly applicable to TMD materials and can be obtained by replacing $\phi_{\rm d, \bar{d}}\to \phi_{\pm 1}$ and $\phi_{\pm 1}\to \phi_{0, \bar 0}$ in (\ref{deltaG}).
After integrating over the fermionic fields, expanding ${\rm Tr ln}\mathcal{G}^{-1}$ up to the second order in $\delta \mathcal{G}^{-1}$ and performing the saddle-point approximation, we obtain the relation between $\phi_{\sigma}$ and $\Delta_{\sigma}$ which is the same as for the previously considered case, but containing spin indices:
\begin{equation}
        \phi_{\sigma}({\bf k}_1,{\bf k}_2,\Omega) = \frac{\Delta_{\sigma}({\bf k}_1,{\bf k}_2,\Omega)}{i[\mathcal{A}_{\sigma}]_{{\bf k}_1{\bf k}_2}^{\,\,\Omega}}, \quad \bar{\phi}_{\sigma}({\bf k}_1,{\bf k}_2,\Omega) = \frac{\bar{\Delta}_{\sigma}({\bf k}_1,{\bf k}_2,\Omega)}{i[\mathcal{A}_{\sigma}]_{{\bf k}_1{\bf k}_2}^{\,\,\Omega}},
\end{equation}
where at $T=0$
\begin{equation}\label{A_sigma}
    \frac{1}{[\mathcal{A}_{\sigma}]_{{\bf k}_1{\bf k}_2}^{\,\,\Omega}} = \begin{cases}
        -i \Omega+ \varepsilon_{c\uparrow}({\bf k}_1) - \varepsilon_{v\uparrow}({\bf k}_2) \quad \text{if}\; \sigma = {\rm d} \\
         -i \Omega +\varepsilon_{c\downarrow}({\bf k}_1) - \varepsilon_{v\uparrow}({\bf k}_2) \quad \text{if}\; \sigma = +1 \\
         -i \Omega +\varepsilon_{c\uparrow}({\bf k}_1) - \varepsilon_{v\downarrow}({\bf k}_2) \quad \text{if}\; \sigma = -1 \\
          -i \Omega +\varepsilon_{c\downarrow}({\bf k}_1) - \varepsilon_{v\downarrow}({\bf k}_2) \quad \text{if}\; \sigma = \bar{\rm d}.
    \end{cases}
\end{equation}
In principle, the expressions in the right-hand side of Eq.~(\ref{A_sigma}) are the exciton dispersions. The chemical potentials of the exciton fields with different spin projections read: $\mu_{c\uparrow} - \mu_{v \uparrow} = \mu_{\rm d}$, $\mu_{c\downarrow} - \mu_{v \uparrow}=\mu_{+1}$, $\mu_{c\uparrow} - \mu_{v \downarrow} =\mu_{-1}$, $\mu_{c\downarrow} - \mu_{v \downarrow} = \mu_{\bar{\rm d}}$. Moreover, under the assumption of the thermodynamic equilibrium between photons and excitons, one gets $\mu_{+1} = \mu_{\rm ph \uparrow}$, $\mu_{-1} =\mu_{\rm ph \downarrow}$. Then the saddle-point approximation for exciton fields yields:
\begin{subequations}\label{saddle-point-eqs-spins}
\begin{eqnarray}
    & \displaystyle \frac{\Delta_{\pm 1}({\bf p},{\bf k}, \Omega)}{[\mathcal{A}_{\pm1}]_{{\bf k}_1{\bf k}_2}^{\,\,\Omega}} & + \frac{g_{\rm R}}{\sqrt{S}}\Psi_{\rm ph, \pm 1}({\bf k}, \Omega) -\sum_{{\bf q}}V({\bf p} -{\bf q}) \Delta_{\pm 1}( {\bf q},{\bf k},\Omega)=0, \\
    & \displaystyle  \frac{\Delta_{\rm d(\bar{\rm d})}({\bf p},{\bf k}, \Omega)}{[\mathcal{A}_{\rm d(\bar{\rm d})}]_{{\bf k}_1{\bf k}_2}^{\,\,\Omega}} & - \sum_{{\bf q}}V({\bf p} -{\bf q}) \Delta_{\rm d(\bar{\rm d})}( {\bf q},{\bf k},\Omega)=0,
\end{eqnarray}
\end{subequations} where ${\bf p}$ and ${\bf k}$ are the relative motion of electrons and the exciton c.m. momenta, respectively. In the c.m. frame the second-order loop $\mathcal{A}^{-1}$ takes the form:
$$\frac{1}{[\mathcal{A}_{\sigma}]_{{\bf k}_1{\bf k}_2}^{\,\,\Omega}} = E_{\rm g} -\mu_{\sigma} + \frac{\hbar^2 {\bf p}^2}{2 m_{\sigma}} + \frac{\hbar^2 {\bf k}^2}{2 M_{\sigma}},$$
with $m_{\sigma}$ being the reduced mass of the $\sigma$-exciton field and $M_{\sigma}$ the $\sigma$-exciton mass.

In the rigid exciton limit the variables can be divided $\Delta_{\sigma}({\bf p}, {\bf k}, \Omega) = \chi_{\sigma}({\bf p})C_{\sigma}({\bf k}, \Omega)$, similarly to the spinless case. The $1s$--state wavefunctions $\chi_{\sigma}({\bf p})$ for different fields are determined by the band splitting:
\begin{equation}
    \frac{p^2}{2m_{\sigma}} \chi_{\sigma}({\bf p}) - \sum_{{\bf q}}V({\bf p-q})\chi_{\sigma}({\bf q})= E_{\rm 1s}^{\sigma}\chi_{\sigma}({\bf p}).
\end{equation}
If the bands for different spins coincide, the expressions are trivial since $\mathcal{A}$ becomes independent of spin.

In principle, the second-order expansion does not deviate significantly from the spinless case and all the loop calculations are straightforward and similar to those presented in Appendix~\ref{appE}, albeit cumbersome. Here we present only the final result.
For the fourth-order term in the expansion series $(1/4){\rm Tr}(\mathcal{G}_0 \delta\mathcal{G}^{-1})^4$
one has 32 loops. Four 
of them describe bright-exciton interactions (two for $|\phi_{+1}|^4$ and two for $|\phi_{-1}|^4$):
\begin{multline}
     \frac{1}{4}{\rm Tr}(\mathcal{G}_{0}\delta\mathcal{G}^{-1})^4 =\frac{2}{4}(i\sqrt{T})^4 \!\!\! \sum_{{\bf k}_1\dots{\bf k}_4}\sum_{\Omega_1\dots\Omega_3} \!\!\! \phi_{\pm 1}({\bf k}_1,{\bf k}_2, \Omega_1)\bar{\phi}_{\pm 1}({\bf k}_3, {\bf k}_2, \Omega_2) \phi_{\pm 1}({\bf k}_3, {\bf k}_4, \Omega_3)\bar{\phi}_{\pm 1}({\bf k}_1, {\bf k}_4, \Omega_1 + \Omega_3 -\Omega_2) \\
     \times \sum_{\omega^\prime}
    \mathcal{G}_{0}({\bf k}_1, \omega^\prime)_{22(11)}\mathcal{G}_0({\bf k}_2, \omega^\prime \! - \Omega_1)_{33(44)}\mathcal{G}_0({\bf k}_3, \omega^\prime \!- \Omega_1+ \Omega_2)_{22(11)}
   \mathcal{G}_0({\bf k}_4, \omega^\prime \!- \Omega_1+ \Omega_2- \Omega_3)_{33(44)}
   \\
   = \frac{1}{4}T \!\! \sum_{{\bf k}_1\dots{\bf k}_4}\sum_{\Omega_1\dots\Omega_3} \!\!\! \Delta_{\pm 1}({\bf k}_1,{\bf k}_2, \Omega_1)\bar{\Delta}_{\pm 1}({\bf k}_3, {\bf k}_2, \Omega_2) \Delta_{\pm 1}({\bf k}_3, {\bf k}_4, \Omega_3)\bar{\Delta}_{\pm 1}({\bf k}_1, {\bf k}_4, \Omega_1 + \Omega_3 -\Omega_2) \times
   \\[4pt]
    \bigl[\!-i \Omega_1 + \varepsilon_{\!c\downarrow(\uparrow)}\!({\bf k}_1\!) - \varepsilon_{\!v\uparrow(\downarrow)}\!({\bf k}_2\!) - i \Omega_3 + \varepsilon_{\!c\downarrow(\uparrow)}\!({\bf k}_3\!) - \varepsilon_{\!v\uparrow(\downarrow)}\!({\bf k}_4) - i \Omega_2 + \varepsilon_{\!c\downarrow(\uparrow)}\!({\bf k}_3\!) - \varepsilon_{\!v\uparrow(\downarrow)}\!({\bf k}_2\!) 
   - i \Omega_1 + i\Omega_2 - i \Omega_3 + \varepsilon_{\!c\downarrow(\uparrow)}\!({\bf k}_1\!) - \varepsilon_{\!v\uparrow(\downarrow)}\!({\bf k}_4)\bigr]
   \nonumber
\end{multline}
which should be treated according to the saddle-point equation leading to the exciton-exciton interaction and saturation terms, similarly with the formulae from Appendix \ref{appB} (it is worth noting that the obtained expression has symmetric form, for convenience).
Dark-exciton interactions $|\phi_{\rm d,\bar{d}}|^4$ contribution has generally the same symmetrized shape (4 diagrams):
\begin{multline}
      \frac{2}{4}T \!\!\! \sum_{{\bf k}_1\dots{\bf k}_4}\sum_{\Omega_1\dots\Omega_3} \!\!\! \Delta_{\rm d,\bar{d}}({\bf k}_1,{\bf k}_2, \Omega_1)\bar{\Delta}_{\rm d,\bar{d}}({\bf k}_3, {\bf k}_2, \Omega_2) \Delta_{\rm d,\bar{d}}({\bf k}_3, {\bf k}_4, \Omega_3)\bar{\Delta}_{\rm d,\bar{d}}({\bf k}_1, {\bf k}_4, \Omega_1 + \Omega_3 -\Omega_2) \\
      \times \bigl[-i \Omega_1 + \varepsilon_{c\uparrow(\downarrow)}({\bf k}_1) - \varepsilon_{v\uparrow(\downarrow)}({\bf k}_2)  -i \Omega_2 + \varepsilon_{c\uparrow(\downarrow)}({\bf k}_3) - \varepsilon_{v\uparrow(\downarrow)}({\bf k}_2)\bigr].
      \nonumber
\end{multline}
Due to the form of the saddle-point equation, this contribution is not leading to saturation,  which is anticipated since the dark excitons are not coupled to photons.

Interestingly, bright-bright and dark-dark interactions do not involve excitons with opposite spins, as long as the Hamiltonian does not 
contain the spin-orbit coupling term [i.e. the Green's function $\mathcal{G}^{-1}_0$ needs to contain non-zero $(\mathcal{G}^{-1}_0)_{13}$ or(and)  $(\mathcal{G}^{-1}_0)_{24}$ elements]. In existing literature~\cite{glazov, wouters_biex, parish_pol} such interactions are also attributed to the biexciton resonance (excitons with opposite spins forming a biecxiton), which cannot be treated within our approach since it does not consider multiple-exciton bound states.

The goal of this Appendix is however to take into account the dark states when considering bright-dark exciton interactions and especially saturation, hence the main subject of our consideration are the terms $|\phi_{+1}|^2 |\phi_{\rm d,\bar{d}}|^2$ and $|\phi_{-1}|^2 |\phi_{\rm d,\bar{d}}|^2$ (16 diagrams: 8 for $+1$ state  and 8 for $-1$ state). All these contributions have generally the same form so we write down only one of them, e.g. $|\phi_{+1}|^2 |\phi_{\rm d}|^2$ (there are 4 loops corresponding to this process):
\begin{multline}
      \frac{2}{4}T \!\! \sum_{{\bf k}_1\dots{\bf k}_4}\sum_{\Omega_1\dots\Omega_3} \!\!\! \bar{\Delta}_{+1}({\bf k}_2,{\bf k}_1, \Omega_1)\Delta_{+1}({\bf k}_2, {\bf k}_3, \Omega_2) \bar{\Delta}_{\rm d}({\bf k}_4, {\bf k}_3, \Omega_3)\Delta_{\rm d}({\bf k}_4, {\bf k}_1, \Omega_1 + \Omega_3 -\Omega_2) \\[4pt]
      \times \bigl[-i \Omega_1 + \varepsilon_{c\downarrow}({\bf k}_2) - \varepsilon_{v\uparrow}({\bf k}_1)  -i \Omega_3 + \varepsilon_{c\uparrow}({\bf k}_4) - \varepsilon_{v\uparrow}({\bf k}_3) -i \Omega_2 + \varepsilon_{c\downarrow}({\bf k}_2) - \varepsilon_{v\uparrow}({\bf k}_3)  - i \Omega_1 +i \Omega_2 -i \Omega_3 + \varepsilon_{c\uparrow}({\bf k}_4) - \varepsilon_{v\uparrow}({\bf k}_1)\bigr].
      \nonumber
\end{multline}
In principle, one should also consider the three other expressions which have the same shape.

Furthermore, the spin-flip processes arise, i.e. $\phi_{+1}\phi_{-1}\bar{\phi}_{\rm d}\bar{\phi}_{\bar{\rm d}}$ and $\phi_{\rm d}\phi_{\bar{\rm d}}\bar{\phi}_{+1}\bar{\phi}_{-1}$ (there are $4+4$ corresponding loops). For example the contribution from $\phi_{\rm d}\phi_{\bar{\rm d}}\bar{\phi}_{+1}\bar{\phi}_{-1}$ in the form symmetrized regarding to the dispersion laws, is as follows:
\begin{multline}
      \frac{2}{4}T \!\! \sum_{{\bf k}_1\dots{\bf k}_4}\sum_{\Omega_1\dots\Omega_3} \!\! \bar{\Delta}_{+1}({\bf k}_2,{\bf k}_1, \Omega_1)\Delta_{\bar{\rm d}}({\bf k}_2, {\bf k}_3, \Omega_2) \bar{\Delta}_{-1}({\bf k}_4, {\bf k}_3, \Omega_3)\Delta_{\rm d}({\bf k}_4, {\bf k}_1, \Omega_1 + \Omega_3 -\Omega_2) \\[4pt]
      \times \bigl[-i \Omega_1 + \varepsilon_{c\downarrow}({\bf k}_2) - \varepsilon_{v\uparrow}({\bf k}_1)  -i \Omega_3 + \varepsilon_{c\uparrow}({\bf k}_4) - \varepsilon_{v\downarrow}({\bf k}_3) -i \Omega_2 + \varepsilon_{c\downarrow}({\bf k}_2) - \varepsilon_{v\downarrow}({\bf k}_3)  - i \Omega_1 +i \Omega_2 -i \Omega_3 + \varepsilon_{c\uparrow}({\bf k}_4) - \varepsilon_{v\uparrow}({\bf k}_1)\bigr].
      \nonumber
\end{multline}

Now we can rewrite the fourth-order contributions in the c.m. variables, taking into account the saddle-point equations, and obtain the exciton-exciton interaction term and the saturation term with corrections due to dark excitons. Even though in the general case the $1s$--exciton wavefunctions are different for $\Delta_{\pm 1,d(\bar d)}$, for simplicity we will assume the band splitting negligible, so the hydrogen-like $1s$--wavefunctions and the effective masses $M_{\sigma}$
will be the same. The effective exciton-photon action becomes $\mathcal{S}_{\rm eff} = \mathcal{S}_0 + \mathcal{S}_{\rm int}^{b} +\mathcal{S}_{\rm int}^d + \mathcal{S}_{\rm int}^{b-d} + \mathcal{S}_{\rm int}^{\rm SF}$, where
\begin{multline}
     \mathcal{S}_0 =  \!\! \int \!\! \frac{d {\bf k}}{(2\pi)^2} \!\! \int \!\frac{d \Omega}{2 \pi} \Biggl\{ \sum_{\substack{\sigma = \pm 1,\\ {\rm d},\bar{\rm d}}} \!\! \overline{C}_{\sigma}({\bf k}, \Omega)\!\left(\!-i\Omega + E_{\rm g} + \frac{\hbar^2 {\bf k}^2}{2 M_{\sigma}} + E_{\rm 1s}^{\sigma} -\mu \! \right) \! C_{\sigma}({\bf k}, \Omega) \Biggr. \\
     \left. + \sum_{\sigma =\pm 1} \!  \left(\overline{\Psi}_{\rm ph, \sigma}({\bf k},\Omega) \bigl[- i \Omega + E_{\rm ph}({\bf k}) -\mu\bigr] \Psi_{\rm ph, \sigma}({\bf k},\Omega) + \frac{\hbar\Omega_{\rm R}}{2} \bigl[\overline{C}_{\sigma}({\bf k}, \Omega)\Psi_{\rm ph, \sigma}({\bf k}, \Omega) +\overline{\Psi}_{\rm ph, \sigma}({\bf k}, \Omega) C_{\sigma}({\bf k}, \Omega) \bigr]\right) \right\}
\end{multline}
is the ``bare'' exciton-photon action without interactions,
\begin{multline}
    \mathcal{S}_{\rm int}^{d} + \mathcal{S}_{\rm int}^{b} =  \!\! \int \! \frac{d{\bf l}_1d{\bf l}_2d{\bf l}_3}{(2\pi)^6} \!\! \int \!\! \frac{d\Omega_1d\Omega_2d\Omega_3}{(2\pi)^3} \Biggl\{\sum_{\substack{\sigma = \pm 1,\\ {\rm d},\bar{\rm d}}} \!\! \frac{g_{\rm ex}}{2} C_{\sigma}({\bf l}_1, \Omega_1) \overline{C}_{\sigma}({\bf l}_2, \Omega_2)  C_{\sigma}({\bf l}_3, \Omega_3) \overline{C}_{\sigma}({\bf l}_1- {\bf l}_2+{\bf l}_3, \Omega_1 -\Omega_2 +\Omega_3) \Biggr.\\
    \left. -  \frac{g_{\rm sat}}{2} \! \sum_{\sigma =\pm 1} \!\!  \Bigl[ C_{\sigma}({\bf l}_1, \Omega_1) \overline{C}_{\sigma}({\bf l}_2, \Omega_2)  C_{\sigma}({\bf l}_3, \Omega_3)\overline{\Psi}_{\rm ph,\sigma}({\bf l}_1- {\bf l}_2+{\bf l}_3, \Omega_1 -\Omega_2 +\Omega_3 ) + \text{c.c.}\Bigr] \right\}
\end{multline}
describe interactions between dark excitons and between bright excitons (within the same species), including bright-exciton-assisted exciton-photon coupling,
\begin{multline}
    \mathcal{S}_{\rm int}^{b-d} \!= \! \int \! \frac{d{\bf l}_1d{\bf l}_2d{\bf l}_3}{(2\pi)^6} \!\! \int \! \frac{d\Omega_1d\Omega_2d\Omega_3}{(2\pi)^3} \! \sum_{\sigma = \pm 1} \!\! \left\{ g_{\rm ex} \Bigl[C_{\rm d}({\bf l}_1, \Omega_1) \overline{C}_{\rm d}({\bf l}_2, \Omega_2) + C_{\bar{\rm d}}({\bf l}_1, \Omega_1) \overline{C}_{\bar{\rm d}}({\bf l}_2, \Omega_2)\Bigr] \right. \\[-5pt]
    \hspace{250pt} \times  C_{\sigma}({\bf l}_3, \Omega_3) \overline{C}_{\sigma}({\bf l}_1 \!-\! {\bf l}_2 \!+\! {\bf l}_3, \Omega_1 \!-\! \Omega_2 \!+\! \Omega_3)  \\[5pt]
    \left.  - \frac{g_{\rm sat}}{2} \Bigl[C_{\rm d}({\bf l}_1, \Omega_1) \overline{C}_{\rm d}({\bf l}_2, \Omega_2) \!+\! C_{\bar{\rm d}}({\bf l}_1, \Omega_1) \overline{C}_{\bar{\rm d}}({\bf l}_2, \Omega_2)\!\Bigr]\!\Bigl[C_{\sigma}({\bf l}_3, \Omega_3)\overline{\Psi}_{\rm ph,\sigma}({\bf l}_1 \!-\! {\bf l}_2 \!+\! {\bf l}_3, \Omega_1 \!-\! \Omega_2 \!+\! \Omega_3 ) + {\rm c.c.}\Bigr]\right\}
\end{multline}
describes interactions between the dark and bright excitons of the same spin, including dark-exciton-assisted exciton-photon coupling and, finally,
\begin{multline}
    \mathcal{S}_{\rm int}^{\rm SF} = \! \int \! \frac{d{\bf l}_1d{\bf l}_2d{\bf l}_3}{(2\pi)^6} \!\! \int \! \frac{d\Omega_1d\Omega_2d\Omega_3}{(2\pi)^3} \biggl\{ g_{\rm ex}^{\rm SF} \,  \overline{C}_{+1}({\bf l}_1, \Omega_1) C_{\bar{\rm d}}({\bf l}_2, \Omega_2)  \overline{C}_{-1}({\bf l}_3, \Omega_3) C_{\rm d}({\bf l}_1 \!-\! {\bf l}_2 \!+\! {\bf l}_3, \Omega_1 \!-\! \Omega_2 \!+\! \Omega_3) \biggr.
    \\[5pt]
    - \frac{g_{\rm sat}}{2} \, C_{\bar{\rm d}}({\bf l}_2, \Omega_2) C_{\rm d}({\bf l}_1- {\bf l}_2+{\bf l}_3, \Omega_1 -\Omega_2 +\Omega_3) \Bigl[\overline{C}_{+1}({\bf l}_1, \Omega_1) \overline{\Psi}_{\rm ph,-1}({\bf l}_3, \Omega_3) + \overline{\Psi}_{\rm ph, +1}({\bf l}_1, \Omega_1) \overline{C}_{-1}({\bf l}_3, \Omega_3)\Bigr] \\[5pt]
    \hspace{95pt} +  g_{\rm ex}^{\rm SF} \, \overline{C}_{\rm d}({\bf l}_1, \Omega_1) C_{-1}({\bf l}_2, \Omega_2)  \overline{C}_{\bar{\rm d}}({\bf l}_3, \Omega_3) C_{+1}({\bf l}_1 \!-\! {\bf l}_2 \!+\! {\bf l}_3, \Omega_1 \!-\! \Omega_2 \!+\! \Omega_3)
    \\[5pt]
    -  \frac{g_{\rm sat}}{2} \, \overline{C}_{\rm d}({\bf l}_1, \Omega_1) \overline{C}_{\bar{\rm d}}({\bf l}_3, \Omega_3) \! \Bigl[\Psi_{\rm ph, \!-1}({\bf l}_2, \Omega_2) C_{\!+1}({\bf l}_1 \!-\! {\bf l}_2 \!+\! {\bf l}_3, \Omega_1 \!-\! \Omega_2 \!+\! \Omega_3) \biggl. +  C_{\!-1}({\bf l}_2, \Omega_2) \Psi_{\rm ph, \!+1}({\bf l}_1 \!-\! {\bf l}_2 \!+\! {\bf l}_3, \Omega_1 \!-\! \Omega_2 \!+\! \Omega_3)\!\Bigr] \!\biggr\}
\end{multline}
corresponds to the spin-flip processes including also the dark-exciton-assisted decay.

It is important to note that other channels of pairing can also be taken into account in the same manner as done in Appendix~\ref{appA} where it was shown that only the {\it off-diagonal} density-like fields of the form $\Phi_{c(v), \sigma,\sigma^\prime}(k_1,k_2)= \overline{\Psi}_{\!c(v), \sigma}(k_1) \Psi_{\!c(v), \sigma^\prime}(k_2)$  contribute to the resulting exciton-exciton interaction (with the inclusion of spins there will be 8 off-diagonal density fields, 4 for conductance and 4 for the valence band). At the same time, the treatment of the 8 remaining {\it diagonal} density-like fields alters the Bethe-Salpeter equation (here, for different exciton fields, which is clearly seen from writing down the third-order expansion terms). When taking into account the additional off-diagonal fields, one needs to calculate 48 vertices of the same type: 24 of those lead to the interaction renormalization of the exciton fields within the same species, the other 24 diagrams screen the interaction between the dark and bright states. Importantly, the spin-flip process is not influenced by the off-diagonal density-like fields.
Thus in all the expressions obtained above for the effective action in the considered simplest case of zero temperature, rigid $1s$--excitons and negligible band splitting, all the bright-bright, dark-dark, and bright-dark interaction constants are renormalized as derived in Eq.~(\ref{gex}).
For spin-flip interactions, however, $g_{\rm ex}$ remains unrenormalized, i.e.
$$g_{\rm ex}^{\rm SF} = 2\int \frac{d{\bf p}}{(2\pi)^2}  \int \frac{d{\bf q}}{(2\pi)^2} V({\bf q})\chi({\bf p})\chi^{*}({\bf p})\chi^{*}({\bf p} -{\bf q})\chi({\bf p} ).$$
 \vor{It is worth noting that in the general case, the $1s$--exciton wavefunctions $\chi_{\sigma}({\bf p})$ corresponding to the fields $\Delta_\sigma$ with different spin indices $\sigma = {\rm b},\bar{\rm b},\textrm{d},\bar{\rm d}$ are different (here $\rm b$, $\bar{\rm b}$ denote the exciton fields with spin projections $1$, $-1$). Therefore the interaction constants describing the interactions and saturation arising between different species change: one would have eight different $g_{\rm ex}$ and eight different $g_{\rm sat}$ constants corresponding to $b - b$, $\bar b-\bar b$, $d - d$, $\bar{d} - \bar{d}$, $b - d(\bar{d})$, $\bar b - d(\bar{d})$ processes, respectively. For the spin-flip processes, the exciton interaction constant $g_{\rm ex}^{\rm SF}$ remains to be of the only value, while the saturation would be described by two different $g_{\rm sat}^{\rm SF}$ corresponding to the exciton-photon conversion with $b$, $\bar b$ spin projections.}

To address corrections to the Rabi-splitting due interactions, we assume for simplicity that photons have polarization $+1$, and follow the procedure as described in Appendix~\ref{appC}, focusing on the $+1$ excitons:
\begin{multline}
     \left. \frac{\delta \mathcal{S}_{\rm eff}}{\delta \overline{C}_{+1}}\right|_{\substack{C_{+1}=\Psi_{\rm ex}\quad\\
     \Psi_{\rm ph, +1} \!= \Psi_{\rm ph} \\  C_{-1}\qquad\quad\\
     C_{\rm d,\bar{d}}\qquad\quad\\
     \Psi_{\rm ph, -1} = 0\quad}} \!\!\!\!\!\! = \!\left(\!E_{\rm g} \!+\! E_{1s}^{(+1)}\! \!-\!\mu\right) \!\Psi_{\rm ex} + \frac{\hbar\Omega_{\rm R}}{2}\Psi_{\rm ph} +g_{\rm ex} |\Psi_{\rm ex}|^2 \Psi_{\rm ex} -g_{\rm sat} |\Psi_{\rm ex}|^2 \Psi_{\rm ph} - \frac{g_{\rm sat}}{2}\Psi_{\rm ex}^2 \overline{\Psi}_{\rm ph}+ g_{\rm ex} (\overline{C}_{\bar{\rm d}}C_{\bar{\rm d}} + \overline{C}_{\rm d}C_{\rm d}) \Psi_{\rm ex} \\[-20pt]
      - \frac{g_{\rm sat}}{2}(\overline{C}_{\bar{\rm d}}C_{\bar{\rm d}} + \overline{C}_{\rm d}C_{\rm d})\Psi_{\rm ph} +
     g_{\rm ex}^{\rm SF} \, \overline{C}_{\!-1}C_{\rm d}C_{\bar{\rm d}} -\frac{g_{\rm sat}}{2} C_{\rm d}C_{\bar{\rm d}}\overline{\Psi}_{\rm ph,-1}=0,
\end{multline}
where the last term is equal zero since only the $+1$ polarization of photons is populated. Analogously,
\begin{multline}
     \left.\frac{\delta \mathcal{S}_{\rm eff}}{\delta\overline{\Psi}_{\rm ph,+1}} \right|_{\substack{
     C_{+1}=\Psi_{\rm ex}\quad \\
     \Psi_{\rm ph, +1} = \Psi_{\rm ph} \\  C_{-1} \qquad\quad\\
     C_{\rm d,\bar{d}} \qquad\quad\\
     \Psi_{\rm ph, -1} = 0\quad}} \!\!\!\!\!\! = (E_{\rm ph}^0-\mu ) \Psi_{\rm ph} +\frac{\hbar\Omega_{\rm R}}{2} \Psi_{\rm ex} - \frac{g_{\rm sat}}{2} |\Psi_{\rm ex}|^2 \Psi_{\rm ex} - \frac{g_{\rm sat}}{2}(\overline{C}_{\bar{\rm d}}C_{\bar{\rm d}} + \overline{C}_{\rm d}C_{\rm d})\Psi_{\rm ex} - \frac{g_{\rm sat}}{2} C_{\rm d}C_{\bar{\rm d}}\overline{C}_{\!-1}=0\,.
\end{multline}
The $(-1)$ exciton branch is assumed to be weakly populated compared to $\Psi_{\rm ex}$, so the terms containing $C_{-1}$ are neglected for clarity. We note that the last term describes the effective decay.

From the above equations, introducing the dark exciton density as $n_{\rm d} = \overline{C}_{\bar{\rm d}} C_{\bar{\rm d}}+ \overline{C}_{\rm d} C_{\rm d}$, we express the bottoms of the UP and LP dispersion laws similarly as for the spinless case:
\begin{equation}\label{shifts_dark}
     \mu_{\rm LP(UP)}(n_{\rm ex}^0, n_{\rm d}) = E_{\rm g} +E_{1s}^{(+1)} + g_{\rm ex} (n_{\rm ex}^0+ n_{\rm d}) +  \frac{\Delta(n_{\rm ex}^0, n_d)}{2} \pm \frac{1}{2}\sqrt{\Delta(n_{\rm ex}^0, n_{\rm d})^2 + [\hbar \Omega_{\rm R}(n_{\rm ex}^0, n_{\rm d})]^2},
\end{equation}
where $\Delta(n_{\rm ex}^0, n_{\rm d}) =\Delta - g_{\rm ex}(n_{\rm ex}^0 + n_{\rm d})$ and $\hbar \Omega_{\rm R}(n_{\rm ex}^0, n_{\rm d}) = \sqrt{[\hbar \Omega_{\rm R} - g_{\rm sat}(n_{\rm ex}^0+ n_{\rm d})][\hbar \Omega_{\rm R} - 3 g_{\rm sat} n_{\rm ex}^0- g_{\rm sat} n_{\rm d}]}$. \vor{It is important to note that Eq.~(\ref{shifts_dark}) contains both the shifts of the two polariton branches do to interaction with dark excitons (of the same size for LP and UP) and due to saturation (with opposite signs).}

The splitting between the polariton branches at the anticrossing point is
\begin{equation}
  \min[E_{\rm UP}({\bf k})- E_{\rm LP}({\bf k})] = \min \sqrt{\Delta_{\bf k}^2(n_{\rm ex}^0, n_{\rm d}) + [\hbar\Omega_{\rm R}(n_{\rm ex}^0, n_{\rm d})]^2},
\end{equation}
with $\Delta_{\bf k}(n_{\rm ex}^0, n_{\rm d}) = \Delta - g_{\rm ex}(n_{\rm ex}^0 +n_{\rm d})+ \hbar^2 {\bf k}^2/2 m_{\rm ph} - \hbar^2 {\bf k}^2/2 M$. As one can see, the presence of dark exciton states effectively reduces the detuning and the Rabi-splitting with the increase of the dark-exciton density. 
\end{widetext}


\begin{thebibliography}{00}
\bibitem{NatRevQuantum} A. Kavokin, T. C. H. Liew, C. Schneider, P. G. Lagoudakis, S. Klembt, and S. Hoefling, {\it Polariton condensates for classical and quantum computing}, \href{https://doi.org/10.1038/s42254-022-00447-1}{Nat Rev Phys {\bf 4}, 435 (2022)}

\bibitem{QFL} I. Carusotto and C. Ciuti, {\it Quantum fluids of light}, \href{https://link.aps.org/doi/10.1103/RevModPhys.85.299}{Rev. Mod. Phys. {\bf 85}, 299 (2013)}

\bibitem{estrecho} E. Estrecho, T. Gao, N. Bobrovska, D. Comber-Todd, M. D. Fraser, M. Steger, K. West, L. N. Pfeiffer, J. Levinsen, M. M. Parish, T. C. H. Liew, M. Matuszewski, D. W. Snoke, A. G. Truscott, and E. A. Ostrovskaya, {\it Direct measurement of polariton-polariton interaction strength in the Thomas-Fermi regime of exciton-polariton condensation}, \href{https://link.aps.org/doi/10.1103/PhysRevB.100.035306}{Phys. Rev. B {\bf 100}, 035306 (2019)}

\bibitem{richard} I. Frérot, A. Vashisht, M. Morassi, A. Lemaître, S. Ravets, J. Bloch, A. Minguzzi, and M. Richard, {\it Bogoliubov excitations driven by thermal lattice phonons in a quantum fluid of light}, \href{https://doi.org/10.1103/PhysRevX.13.041058}{Phys. Rev. X {\bf 13}, 041058 (2023)}

\bibitem{menon}  J. Gu, V. Walther, L. Waldecker, D. Rhodes, A. Raja, J. C. Hone, T. F. Heinz, S. Kéna-Cohen, T. Pohl, and V. M. Menon, {\it Enhanced nonlinear interaction of polaritons via excitonic Rydberg states in monolayer WSe$_2$}, \href{https://doi.org/10.1038/s41467-021-22537-x}{Nat Commun {\bf 12}, 2269 (2021)}

\bibitem{zhao} J. Zhao, A. Fieramosca, K. Dini, R. Bao, W. Du, R. Su, Y. Luo, W. Zhao, D. Sanvitto, T. C. H. Liew, and Q. Xiong, {\it Exciton polariton interactions in Van der Waals superlattices at room temperature}, \href{https://doi.org/10.1038/s41467-023-36912-3}{Nat. Commun {\bf 14}, 1512 (2023)}

\bibitem{stepanov} P. Stepanov, A. Vashisht, M. Klaas, N. Lundt, S. Tongay, M. Blei, S. Höfling, T. Volz, A. Minguzzi, J. Renard, C. Schneider, and M. Richard, {\it Exciton-exciton interaction beyond the hydrogenic picture in a MoSe$_2$ monolayer in the strong light-matter coupling regime}, \href{https://doi.org/10.1103/PhysRevLett.126.167401}{Phys. Rev. Lett. {\bf 126}, 167401 (2021)}

\bibitem{deng_const} L. Zhang, F. Wu, Sh. Hou, Zh. Zhang, Y.-H. Chou, K. Watanabe, T. Taniguchi, S. R. Forrest, and H. Deng, {\it Van der Waals heterostructure polaritons with moiré-induced nonlinearity}, \href{https://doi.org/10.1038/s41586-021-03228-5}{Nature {\bf 591}, 61–65 (2021)}

\bibitem{barachati} F. Barachati, A. Fieramosca, S. Hafezian, J. Gu, B. Chakraborty, D. Ballarini, L. Martinu, V. Menon, D. Sanvitto, and St. Kéna-Cohen, {\it  Interacting polariton fluids in a monolayer of tungsten disulfide}, \href{https://doi.org/10.1038/s41565-018-0219-7}{Nat. Nanotechnol. 13, 906 (2018)}


\bibitem{BEC_polaritons} J. Kasprzak, M. Richard, S. Kundermann, A. Baas, P. Jeambrun, J. M. J. Keeling, F. M. Marchetti, M. H. Szymańska, R. André, J. L. Staehli, V. Savona, P. B. Littlewood, B. Deveaud, and Le Si Dang, {\it Bose–Einstein condensation of
exciton polaritons}, \href{https://doi.org/10.1038/nature05131}{Nature {\bf 443}, 409 (2006)}

\bibitem{SF_polaritons} A. Amo, J. Lefrère, S. Pigeon, C. Adrados, C. Ciuti, I. Carusotto, R. Houdré, E. Giacobino, and A. Bramati, {\it Superfluidity of polaritons in semiconductor microcavities}, \href{https://doi.org/10.1038/nphys1364}{Nature Phys {\bf 5}, 805–810 (2009)}

\bibitem{yamamoto} F. Tassone and Y. Yamamoto, {\it Exciton-exciton scattering dynamics in a semiconductor microcavity and stimulated scattering into polaritons}, \href{https://doi.org/10.1103/PhysRevB.59.10830}{Phys. Rev. B {\bf59}, 10830 (1999)}

\bibitem{glazov} M. M. Glazov, H. Ouerdane, L. Pilozzi, G. Malpuech, A. V. Kavokin, and A. D’Andrea, {\it Polariton-polariton scattering in microcavities: A microscopic theory}, \href{https://doi.org/10.1103/PhysRevB.80.155306}{Phys. Rev. B {\bf80}, 155306 (2009)}

\bibitem{combescot} M. Combescot, O. Betbeder-Matibet, and F. Dubin, {\it The many-body physics of composite bosons}, \href{https://doi.org/10.1016/j.physrep.2007.11.003}{Phys. Rep. {\bf 463}, 215 (2008)}

\bibitem{parish_ehph1} J. Levinsen, G. Li, and M. M. Parish, {\it Microscopic description of exciton-polaritons in microcavities}, \href{https://doi.org/10.1103/PhysRevResearch.1.033120}{Phys. Rev. Research {\bf 1}, 033120 (2019)}

\bibitem{schwendimann}G. Rochat, C. Ciuti, V. Savona, C. Piermarocchi, A. Quattropani, and P. Schwendimann, {\it Excitonic Bloch equations for a two-dimensional system of interacting excitons}, \href{https://doi.org/10.1103/PhysRevB.61.13856}{ Phys. Rev. B {\bf 61}, 13856 (2000)}

\bibitem{binder_scat} R. Takayama, N.H. Kwong, I. Rumyantsev, M. Kuwata-Gonokami, and R. Binder, {\it T-matrix analysis of biexcitonic correlations in the nonlinear optical response of semiconductor quantum wells}, \href{https://doi.org/10.1140/epjb/e20020051}{Eur. Phys. J. B {\bf 25}, 445 (2002)}

\bibitem{kira} M. Kira, F. Jahnke, S. W. Koch, J. D. Berger, D. V. Wick, T. R. Nelson, Jr., G. Khitrova, and H. M. Gibbs, {\it Quantum Theory of Nonlinear Semiconductor Microcavity Luminescence Explaining “Boser” Experiments}, \href{https://doi.org/10.1103/PhysRevLett.79.5170}{Phys. Rev. Lett. {\bf 79}, 5170 (1997)}

\bibitem{khurgin1} D. S. Citrin and J. B. Khurgin, {\it Microcavity effect on the electron-hole relative motion in semiconductor quantum wells}, \href{https://doi.org/10.1103/PhysRevB.68.205325}{Phys. Rev. B {\bf 68}, 205325 (2003)}



\bibitem{PI4} P. R. Eastham and P. B. Littlewood, {\it Bose condensation of cavity polaritons beyond the linear regime: The thermal equilibrium of a model microcavity}, \href{https://doi.org/10.1103/PhysRevB.64.235101}{Phys. Rev. B {\bf64}, 235101 (2001)}

\bibitem{PI1} J. Keeling, P. R. Eastham, M. H. Szymanska, and P. B. Littlewood, {\it Polariton condensation with localized excitons and propagating photons}, \href{https://doi.org/10.1103/PhysRevLett.93.226403}{Phys. Rev. Lett. {\bf93}, 226403 (2004)}

\bibitem{PI3} M. H. Szymańska, J. Keeling, and P. B. Littlewood, {\it Nonequilibrium quantum condensation in an incoherently pumped dissipative system}, \href{https://doi.org/10.1103/PhysRevLett.96.230602}{Phys. Rev. Lett. {\bf 96}, 230602 (2006)}

\bibitem{PI5} F. M. Marchetti, J. Keeling, M. H. Szymańska, and P. B. Littlewood, {\it Absorption, photoluminescence, and resonant Rayleigh scattering probes of condensed microcavity polaritons}, \href{https://doi.org/10.1103/PhysRevB.76.115326}{Phys. Rev. B {\bf76}, 115326 (2007)}

\bibitem{PI2} M. H. Szymańska, J. Keeling, and P. B. Littlewood, {\it Mean-field theory and fluctuation spectrum of a pumped decaying Bose-Fermi system across the quantum condensation transition}, \href{https://doi.org/10.1103/PhysRevB.75.195331}{Phys. Rev. B {\bf75}, 195331 (2007)}

\bibitem{PI6} J. Keeling, P. R. Eastham, M. H. Szymanska, and P. B. Littlewood, {\it BCS-BEC crossover in a system of microcavity polaritons}, \href{https://doi.org/10.1103/PhysRevB.72.115320}{Phys. Rev. B {\bf 72}, 115320 (2005)}

\bibitem{parish_ehph3} S. S. Kumar, M. M. Parish, and J. Levinsen, {\it Microscopic theory of excitons bound by light}, \href{https://doi.org/10.1103/PhysRevB.106.205414}{Phys. Rev. B {\bf 106}, 205414 (2022)}

\bibitem{keldysh} L.V. Keldysh, A.N. Kozlov, {\it Collective Properties of Excitons in Semiconductors}, JETP {\bf 54}, 978 (1968)

\bibitem{kiselev} V.S. Babichenko and M.N. Kiselev, {\it Superconductivoty in systems with excitonic instability}, Journal of Moscow Phys. Soc. {\bf 2}, 311 (1992)

\bibitem{tokatly} A. A. Gorbatsevich and I. V. Tokatly, {\it Phase transitions in an exciton semiconductor and in a superfluid low-density Fermi liquid: Bose-liquid approach. Three and two-dimensional systems. General description at finite temperatures},  JETP {\bf 108}, 1723 (1995)

\bibitem{berkelbach_prb88} T. C. Berkelbach, M. S. Hybertsen, and D. R. Reichman, {\it Theory of neutral and charged excitons in monolayer transition metal dichalcogenides}, \href{https://link.aps.org/doi/10.1103/PhysRevB.88.045318}{Phys. Rev. B {\bf 88}, 045318 (2013)}

\bibitem{chernikov_prl113} A. Chernikov, T. C. Berkelbach, H. M. Hill, A. Rigosi, Y. Li, B. Aslan, D. R. Reichman, M. S. Hybertsen, and T. F. Heinz, {\it Exciton Binding Energy and Nonhydrogenic Rydberg Series in Monolayer
WS$_2$}, \href{https://link.aps.org/doi/10.1103/PhysRevLett.113.076802}{Phys. Rev. Lett. {\bf 113}, 076802 (2014)}

\bibitem{keldysh_int} L. V. Keldysh, {\it Coulomb interaction in thin semiconductor and semimetal films}, JETP Lett. {\bf 29}, 658 (1979)

\bibitem{parish_pol} O. Bleu, G. Li, J. Levinsen, and M. M. Parish, {\it Polariton interactions in microcavities with atomically thin semiconductor layers}, \href{https://doi.org/10.1103/PhysRevResearch.2.043185}{Phys. Rev. Research {\bf 2}, 043185 (2020)}

\bibitem{LozYudson} Yu. E. Lozovik and V. I. Yudson, {\it On the ground state of the two-dimensional non-ideal bose gas}, \href{https://doi.org/10.1016/0378-4371(78)90170-X}{Physica A {\bf 93}, 493 (1978)}

\bibitem{Castin_notes} Y. Castin, {\it Simple theoretical tools for low dimensional Bose gases}, \href{https://doi.org/10.1051/jp4:2004116004}{J. Phys. IV {\bf 116}, 89 (2004)}

\bibitem{trion-polaritons} R. P. A. Emmanuele, M. Sich, O. Kyriienko, V. Shahnazaryan, F. Withers, A. Catanzaro, P. M. Walker, F. A. Benimetskiy, M. S. Skolnick, A. I. Tartakovskii, I. A. Shelykh, and D. N. Krizhanovskii, {\it Highly nonlinear trion-polaritons in a monolayer semiconductor}, \href{https://doi.org/10.1038/s41467-020-17340-z}{Nat Commun {\bf 11}, 3589 (2020)}

\bibitem{keldysh_1} L.V. Keldysh and A.N. Kopaev, {\it  Possible instability of the semimetallic state toward Coulomb interaction}, Soviet Physics Solid State {\bf 6}, 2219 (1965)

\bibitem{marchetti} F. M. Marchetti, B. D. Simons, and P. B. Littlewood, {\it Condensation of cavity polaritons in a disordered environment}, \href{https://doi.org/10.1103/PhysRevB.70.155327}{Phys. Rev. B {\bf70}, 155327 (2004)}

\bibitem{lozovik_PI} A. Elistratov, Yu. Lozovik, {\it Coupled exciton-photon Bose condensate in path integral formalism}, \href{https://doi.org/10.1103/PhysRevB.93.104530}{Phys. Rev. B {\bf93}, 104530 (2016)}

\bibitem{khurgin} J.B. Khurgin, {\it Excitonic radius in the cavity polariton in the regime of very strong coupling}, \href{https://doi.org/10.1016/S0038-1098(00)00469-5}{Solid State Communications, {\bf 117}, 307 (2001)}

\bibitem{Shahnazaryan2017} V. Shahnazaryan, I. Iorsh, I. A. Shelykh, and O. Kyriienko, {\it Exciton-exciton interaction in transition-metal dichalcogenide monolayers}, \href{https://doi.org/10.1103/PhysRevB.96.115409}{Phys. Rev. B {\bf96}, 115409 (2017)}

\bibitem{numericalWF} For hydrogen-like wavefunctions in TMDs, in order to obtain the values presented in Table~\ref{table_const}, we used $a_{\rm MoSe_2} = 1.1$~nm and $a_{\rm WS_2} = 1.8$~nm (see e.g.~\cite{berkelbach_prb88}). We note that for the accurate wavefunctions (numerically simulated with the Rytova-Keldysh potential), the interaction constants substantially deviate from the hydrogenic approximation. In particular, without assuming the variational hydrogen-like ansatz we obtain for Ref.~\cite{menon} $g_{\rm ex} = 1.18~\mu$eV$\mu$m$^2$, $g_{\rm sat} = 0.14~\mu$eV$\mu$m$^2$; for Ref.~\cite{zhao} $g_{\rm ex} = 1.18~\mu$eV$\mu$m$^2$, $g_{\rm sat} = 0.17~\mu$eV$\mu$m$^2$; for Ref.~\cite{stepanov} $g_{\rm ex} = 0.74~\mu$eV$\mu$m$^2$, $g_{\rm sat} = 0.073~\mu$eV$\mu$m$^2$; and for Ref.~\cite{deng_const} $g_{\rm ex} = 0.75~\mu$eV$\mu$m$^2$, $g_{\rm sat} = 0.057~\mu$eV$\mu$m$^2$.

\bibitem{opala} A. Opala, M. Furman, M. Król, R. Mirek, K. Tyszka, B. Seredyński, W. Pacuski, J. Szczytko, M. Matuszewski, and B. Piętka, {\it Natural exceptional points in the excitation spectrum of a light–matter system}, \href{https://doi.org/10.1364/OPTICA.497170}{Optica {\bf10}, 1111 (2023)}

\bibitem{wouters_biex} M. Wouters, {\it Resonant polariton-polariton scattering in semiconductor microcavities}, \href{https://doi.org/10.1103/PhysRevB.76.045319}{Phys. Rev. B {\bf 76}, 045319 (2007)}

\bibitem{vladimirova} M. Vladimirova, S. Cronenberger, D. Scalbert, K. V. Kavokin, A. Miard, A. Lemaître, J. Bloch, D. Solnyshkov, G. Malpuech, and A. V. Kavokin, {\it Polariton-polariton interaction constants in microcavities}, \href{https://doi.org/10.1103/PhysRevB.82.075301}{Phys. Rev. B {\bf 82}, 075301 (2010)}

\bibitem{parish_biex} K. Choo, O. Bleu, J. Levinsen, and M. M. Parish, {\it Polaronic polariton quasiparticles in a dark excitonic medium}, \href{https://doi.org/10.48550/arXiv.2312.00985}{arXiv:2312.00985}

\bibitem{shelykh_biex} A. Kudlis, I. A. Aleksandrov, M. M. Glazov, and I. A. Shelykh, {\it Theory of biexciton-polaritons in transition metal dichalcogenide monolayers}, \href{https://doi.org/10.48550/arXiv.2402.09110}{arXiv:2402.09110}

\bibitem{HS} R.L. Stratonovich, {\it On a Method of Calculating Quantum Distribution Functions}, Soviet Physics, Doklady, {\bf 2}, 416 (1958); J. Hubbard, {\it Calculation of Partition Functions}, \href{https://doi.org/10.1103/PhysRevLett.3.77}{Phys. Rev. Lett. {\bf 3}, 77 (1959)}

\end{thebibliography}
\end{document}